\documentclass[12pt,a4paper]{article}
\usepackage{graphicx, array, epsfig, subfigure}
\usepackage{graphicx,amssymb}
\usepackage{epstopdf}
\newcommand{\be}{\begin{equation}}
\newcommand{\ee}{\end{equation}}
\newcommand{\bea}{\begin{eqnarray}}
\newcommand{\eea}{\end{eqnarray}}

\begin{document}
\begin{center}
{\bf Two Component Feebly Interacting Massive Particle (FIMP) Dark Matter}\\
\vspace{1cm}
{{\bf Madhurima Pandey} \footnote{email: madhurima.pandey@saha.ac.in}, 
{\bf Debasish Majumdar} \footnote{email: debasish.majumdar@saha.ac.in}}\\
{\normalsize \it Astroparticle Physics and Cosmology Division,}\\
{\normalsize \it Saha Institute of Nuclear Physics, HBNI} \\
{\normalsize \it 1/AF Bidhannagar, Kolkata 700064, India}\\
\vspace{0.25cm}
{\bf Kamakshya Prasad Modak} \footnote{email: kamakshya.modak@gmail.com}\\
{\normalsize \it Department of Physics, Brahmananda Keshab Chandra College,}\\
{\normalsize \it 111/2, B. T. Road, Kolkata 700108, India}\\
\vspace{1cm}
\end{center}
\begin{abstract}
We explore the idea of an alternative candidate for particle dark matter
namely Feebly Interacting Massive Particle (FIMP) in the framework of a
two component singlet scalar model. Singlet scalar dark matter has
already been demonstrated  to be a viable candidate 
for WIMP (Weakly Interacting
Massive Particle) dark matter in literature. In the FIMP scenario,
dark matter particles are slowly produced via ``thermal frreze-in'' mechanism in
the early Universe and
are never abundant enough to reach thermal equilibrium or to undergo
pair annihilation inside the Universe's plasma due to their extremely
small couplings. We demonstrate that for smaller couplings too,
required for freeze-in process, a two component scalar
dark matter model considered here could well be a viable candidate for FIMP.
In this scenario, the Standard Model of particle physics is 
extended by two gauge singlet
real scalars whose stability is protected by an unbroken $Z_{2}\times {Z'}_{2}$
symmetry and they are assumed to acquire no VEV after Spontaneous Symmetry
Breaking. We explore the viable mass regions in the present two scalar DM
model that is in accordance with the FIMP scenario.
We also explore the upper limits of masses of the two components from 
the consideration of their self  interactions.
\end{abstract}

\section{Introduction}
One of the most important problems of fundamental physics is to
ascertain the particle nature of dark matter (DM) and their production
mechanisms in the early Universe. 
The existence of dark matter in the Universe is established only
through its gravitational effects and from different astronomical 
and cosmological observations such as rotation curves of spiral 
galaxies \cite{sofue}, gravitational lensing \cite{barte}, phenomenon 
of Bullet cluster \cite{clowe}, PLANCK \cite{ade} satellite borne
 experiment for measuring the anisotropies 
in Cosmic Microwave Background Radiation (CMBR) etc. The direct 
evidence of dark matter 
through the direct detection mechanism \cite{Akerib:2016vxi}-\cite{Tan:2016zwf}
whereby a detector nucleus scatters off by a possible DM particle is yet to
be found.
One of the viable and 
popular candidates for dark matter may be the WIMPs (Weakly Interacting
Massive Particles) \cite{jungman}-\cite{srednicki}. But the particle 
candidates for WIMPs are not known yet. Also not known whether 
the dark matter in the Universe is made up of  one particle component 
or its constituent components are more than one.

Although the WIMPs are yet to be detected in the world wide endeavour
for direct dark matter search they continue to be popular dark matter 
candidates. WIMPs are produced in the early Universe
thermally and they maintained thermal and chemical equilibrium at that epoch.
When expansion rate of the Universe
exceeded the interaction rates of the DM particles, these particles were not
able to interact with them anymore. As a result such dark matter particles  
fell out of (or moved away from)
equilibrium. Thus they suffered a state of ``freeze out'' by being 
decoupled from the Universe's plasma and remained as relics. 
There are abundant examples in the literature where various particle physics 
models are proposed for viable particle candidates of WIMP dark matter. Such 
models are either based on simple extensions of Standard Model of
particle physics (SM)
or other established theories Beyond Standard Model
(BSM). Some well-known candidates in the latter category are the 
neutralinos \cite{jungman} in supersymmetric (SUSY) theories, the lightest
Kaluza-Klein particle \cite{cheng} in Universal Extra-Dimensional (UED) model,
the singlet \cite{Silveira:1985rk}-\cite{Barger:2007im} etc. while 
the former category includes, among other models, the singlet and doublet
 \cite{Ma:2006km}-\cite{Banik:2014cfa} scalar extensions
of the SM, singlet fermionic dark matter 
\cite{Kim:2008pp}-\cite{Fairbairn:2013uta},
hidden sector vector dark matter 
\cite{Hambye:2008bq}-\cite{matti1} etc. Vector and scalar dark matter in a model with scale invariant SM extended by a dark sector has been explored in \cite{karam,karam1}. Fermion dark matter in a dark sector (with gauge group $\rm SU(2) \times \rm U(1)$) and dark $\rm U(1)$ charge are considered by Biswas et. al. \cite{banik}.
But there is no definitive evidences that dark matter is in fact consists of 
WIMP particles \cite{arcadi} or other particles that are thermally produced in 
the early Universe.
It is important therefore to consider viable
alternatives to thermal WIMPs.

In this work  we explore, for a viable dark matter candidate, a well
motivated alternative to the WIMP mechanism, namely the FIMP
(Feebly Interacting Massive Particle) \cite{McDonald:2001vt,Hall:2009bx,sami,dev}
mechanism. Here, we propose a dark matter candidate that has two components and
the production of which in the early Universe are assumed to be 
through the
FIMP mechanism. FIMPs are identified by their small interaction 
rates with
Standard Model particles in the early Universe. Due to such feeble interactions
these FIMP particles  are unable to
reach thermal equilibrium with the Universe's plasma throughout their
cosmological history. 
FIMPs are thus slowly produced by decays or
annihilations of Standard Model particles in the thermal plasma
and in contrast to WIMPs they are never abundant enough to undergo
annihilation interactions among themselves. 
Therefore they are never in thermal or chemical 
equilibrium with rest of the Universe's plasma. 
But their number densities increase slowly due to their very 
small couplings with the SM particles.
Thus, in contrast to thermal WIMP cases where the dark matter
particles go away from the equilibrium, the FIMP particles approach
towards equilibrium. 
An example of a FIMP candidate
may be sterile neutrino, which is produced from the decay of some
heavy scalars \cite{Merle:2014xpa}-\cite{Shakya:2015xnx} or gauge bosons
\cite{Biswas:2016bfo}.
In Refs. \cite{Yaguna:2011qn}-\cite{Biswas:2016iyh} various FIMP type DM
candidates have been discussed.

In the present work we propose a two component dark matter model in 
FIMP scenario. The model involves two
distinct singlet scalars that serve as the two components of the dark 
matter. Our purpose is to demonstrate the viablity 
of such two 
component singlet scalars in FIMP scenario to be  
dark matter candidates
in the mass regimes spanning from GeV to keV. 
To this end, three pairs of masses are considered for the dark matter
components in the mass regimes GeV, MeV and keV. 
For a one component singlet scalar dark matter model, the Standard 
Model is minimally extended by an additional scalar singlet
\cite{Silveira:1985rk}. 
In our work (involving two scalar components) we extend the 
scalar sector of SM by two real scalar 
fields $S_{2}$ and $S_{3}$, both of which are singlets under 
the Standard Model gauge group $\rm SU(2)_{L} \times \rm U(1)_{Y}$. 
Productions of both dark matter components in FIMP scenario proceed 
from the pair annihilation of SM particles such as fermions, gauge 
bosons and Higgs bosons. These scalars are assumed to acquire no 
vacuum expectations values (VEV) at 
spontaneous symmetry breaking (SSB) and a ${Z}_2 \times {Z'}_2$ 
symmetry \cite{kamakshya,ghosh,ghosh1} is imposed on the two scalars of the extended scalar sector so as to 
prevent the interactions of the two scalar components with 
the SM fermions or their decays.
Here discrete symmetries $Z_2$ and ${Z'}_2$ are imposed on the scalars 
$S_2$ and $S_3$ respectively.
As both the scalars do not generate any VEV at SSB, 
the fermion masses are also not affected. Such a scalar interacts 
with the SM sector only through a Higgs portal due to the interaction 
term (in inteaction Lagrangian) of the type 
${H}^\dagger  H S_{i}S_{i}$  (where $i=2,3$). 
The unknown couplings of these additional scalars are the parameters of 
the theory. These can be constrained 
using the theoretical bounds on the Lagrangian as also by computing 
the relic densities and then comparing them with the
same given by PLANCK experiment. 

The relic densities are calculated by evaluating the
comoving 
number densities $n_{s_2}$ and $n_{s_3}$ (which are in general written 
in terms of the ratios $Y_{s_{2}} = \frac {n_{S_2}} {S}$ and 
$Y_{s_{3}} = \frac {n_{s_3}} {S}$ of
corresponding number densities and the entropy density {$S$}
of the Universe) for the scalar dark matter components $s_2$ and 
$s_{3}$ respectively (see later (Sect. 3)). At the present epoch these are computed by 
solving self consistently, the relevant coupled 
Boltzmann equations for the two components. As mentioned earlier, in FIMP 
scenario the number density of a species evolves towards
its equilibrium value from almost negligible initial 
abundance. This means, initially $Y_{s_{2}} \approx 0 \approx Y_{s_{3}}$. 
Evolution of these abundances requires computations of the 
quantities such as the decay processes 
$h \rightarrow s_js_j$ ($j=2,3$), where $h$ denotes the SM 
Higgs, the pair annihilation processes $ x\bar{x}\rightarrow s_js_j$, 
where $x$ can be $ W^{\pm}, Z, f(\bar { f}),h,s_2,s_3$ etc. 
The total dark matter relic density for the considered two component 
singlet scalar model in FIMP scenario is finally obtained 
by adding the computed 
individual abundances of each of the components as 
$\Omega_{\rm tot} {\tilde {h}}^{2} = \Omega_{s_{2}}{\tilde h}^{2}+
\Omega_{s_{3}}{\tilde {h}}^{2}$,
where the relic density $\Omega$ for a particular species is expressed 
in terms of $\Omega {\tilde {h}}^{2}$, ${\tilde {h}}$ being the Hubble parameter
normalised to 100 km s$^{-1}$ Mpc$^{-1}$. The computed value of 
$\Omega_{\rm tot} {\tilde {h}}^{2}$
should be consistent with WMAP \cite{hinshaw}/PLANCK \cite{ade} observational
results,  $0.1172\leq \Omega_{\rm DM} {\tilde {h}}^{2} \leq 0.1226$.

There are indications from astronomical observations of collisions of 
galaxies and galaxy clusters, the existence
of self interactions among the dark matter paricles. From observations of
Bullet Cluster phenomenon in the past and from more recent observations
of 72 colliding galaxy clusters \cite{Harvey:2015hha}, an upper 
limit to the dark matter self interacting cross-sections per unit 
dark matter mass has been given in the literatures. In this work we 
explore further the mass regions for our two component FIMP dark matter
model that agrees to this self interaction bound while satisfying other
conditions mentioned earlier.   

The paper is organised as follows. In Section 2 we give a brief account of the
Freeze-in process. Section 3 furnishes our two component scalar dark matter 
model whicle Section 4 deals with the constraints by which the model parameter 
space can be constrained. In Section 5 the methodology to compute the relic 
densities for the present two component scalar DM model is described. 
In Section 6 we furnish our calculational results considering 
the dark matter candidates in three mass regions 
namely GeV, MeV and keV. As mentioned, we have 
also computed the dark matter self interactions for the present scenario. The 
formalism computations for the same are given in Section 7. Finally in 
Section 8 we give a brief summary and discussions.   

\section{Freeze-in Overview}

In this section, we briefly discuss the ``freeze-out" mechanism  for 
thermal production of the dark matter. In the early Universe massive DM 
candidates could be thermally produced by the collision of the particles in 
thermal cosmic 
plasma and were in both kinetic and chemical equilibrium with the thermal plasma. 
Dark matter particles have a large initial thermal density at a 
temperature $T$ which 
is greater than the mass of DM ($m_{\chi}$, where $\chi$ denotes 
a thermal dark matter candidate). As the temperature of the hot plasma 
of the early Universe dropped 
below the mass of the dark matter, the lighter particles lacked the 
potential to produce heavier particles as they no more have 
enough kinetic energy (thermal energy). The expansion of the Universe dilutes 
the number of particles and thus interaction between them can hardly occur. 
Thus the conditions for thermal equilibrium were violated. The DM particles 
then go away from the equilibrium and decouple from Universal hot plasma. This
phenomenon is called ``freeze-out''. After ``freeze-out'' the comoving 
number density of DM particles became fixed and these particles remain as 
relic. Larger the annihilation cross-sections of the particles more is the 
annihilation of DM particles before freeze out and consequently the density 
will be less. An attractive feature of the freeze-out 
mechanism is that for renormalisable couplings the yield is dominated by low 
temperatures with freeze-out typically occuring at a temperature which is 
a factor $\sim 20-25$ of the DM mass. The WIMPs are generally produced 
through this mechanism.

As mentioned in Section 1, we explore in this paper, a dark matter 
candidate that is 
produced through an alternate mechanism namely the 
``freeze-in'' mechanism from almost negligible initial abundance and with 
very feeble interactions with other particles. 
The dark matter produced through this 
mechanism is generally referred to as 
Feebly Interacting Massive Particles (FIMPs) dark matter. 
As the interaction of such FIMPs with other bath particles 
(Standard Model particles) are very feeble, they never attain 
thermal equilibrium. Although feeble, initially the FIMPs 
production may happen slowly due to the very feeble interactions with the 
Standard Model particles which grow gradually. The dominant production 
occurs at $T \sim m_{\chi}$, where $T$ is 
the temperature of the Universe. 

The freeze-in process is opposite in nature to that of freeze-out. 
As the temperature $T$ drops below 
the mass of the relevant particle (here the DM candidate), the DM is either 
heading away from 
(freeze-out) or towards (freeze-in) thermal equilibrium. In freeze-out 
mechanism the initial number density varies as $T^{3}$ and then decreases as  
the interaction strength reduces to maintain this large abundance. On the other hand 
freeze-in has a negligible initial DM abundance, and increases as the 
interaction strength   increases the production of DM from the thermal bath.

\section{Two Component Dark Matter Model}

The two component dark matter model in FIMP scenario proposed in this 
work consists of two distinct scalar singlet DM particles $S_2$ and $S_3$. 
Here, we have a renormalisable extension of the SM by adding two 
real scalar fields $S_2$ and $S_3$. These two real scalars are singlets 
under the SM gauge group and they are stabilised by imposing a 
discrete $Z_2 \times Z_2'$ symmetry. These scalars do not generate 
any VEV after spontaneous symmetry breaking and there is no mixing 
between these real scalars and the SM scalar. The only possible way 
that the DM candidates interact with the SM sector is  through Higgs portal.

The Lagrangian of our model can be written as
\bea
{\cal L} = {\cal L}_{\rm SM} + {\cal L}_{\rm DM} + {\cal L}_{\rm int}\,\, ,
\label{lagrangian}
\eea
where $\mathcal{L}_{\rm SM}$ stands for the Lagrangian of the SM particles 
and it consists of quadratic and quartic terms involving the Higgs doublet $H$ 
in addition to the usual kinetic term for $H$.
As mentioned, the dark sector Lagrangian consists of two real scalar fields, 
which can be expressed as
\bea
{\cal L}_{\rm DM} = {\cal L}_{S_2} + {\cal L}_{S_3}\,\, ,
\label{lagrangian1}
\eea
with
\bea
{\cal L}_{S_2} = \frac{1}{2} (\partial_{\mu} S_2)(\partial^{\mu} S_2) -
\frac{{ \mu}_{S_2}^2}{2} S_2^2 - \frac{\lambda_{S_2}}{4} S_2^4\,\, ,
\label{lagS2}
\eea
and 
\bea
{\cal L}_{S_3} = \frac{1}{2} (\partial_{\mu} S_3)(\partial^{\mu} S_3) -
\frac{ { \mu}_{S_3}^2}{2} S_3^2 - \frac{\lambda_{S_3}}{4} S_3^4\,\, .
\label{lagS3}
\eea
The interaction Lagrangian $\mathcal{L}_{\rm int}$ contains all possible
mutual interaction terms among the scalar fields $H, S_2, S_3$.
\bea
{\cal L}_{\rm int} = -V' (H, S_2, S_3)\,\, ,
\label{lagint}
\eea
where $V' (H, S_2, S_3)$ can be written as
\bea
V^{\prime}(H,\,S_2,\,S_3) ={\lambda_{H S_2}} H^\dagger H\, S_2^2
+ {\lambda_{HS_3}} H^\dagger H\,S_3^2
+ \lambda_{S_2 S_3} S_2^2\,S_3^2 \,\, .
\label{int}
\eea
The renormalisable scalar potential $V$ is written as
\bea
V &=& \mu_{H}^2\,H^\dagger H + \lambda_{H}\,(H^\dagger H)^2
+ \frac {\mu_{S_2}^2} {2} S_2^2 + \frac {\lambda_{S_2}}{4} S_2^4
+ \frac{ { \mu}_{S_3}^2}{2} S_3^2 + \frac{\lambda_{S_3}}{4} S_3^4 \nonumber \\
&& + {\lambda_{H S_2}} H^\dagger H\, S_2^2
+ {\lambda_{HS_3}} H^\dagger H\,S_3^2 
+ \lambda_{S_2 S_3} S_2^2\,S_3^2 \,.
\label{potential}
\eea
After the spontaneous symmetry breaking SM Higgs acquires a VEV, 
$v$ ($v$ $\sim$ 246 GeV) and SM scalar doublet takes the form 
\bea
H = \frac {1} {\sqrt {2}} \left ( \begin{array}{c} 0 \\ v + h \end{array}
\right )\,\, . 
\label{higgs}
\eea
It is assumed in the present model that the two scalars $S_2$ and $S_3$ 
do not generate any VEV such that 
$\langle S_2 \rangle = 0 = \langle S_3 \rangle $. As a result, after 
SSB we have $H \rightarrow h + v$, $S_2 = s_2 + 0, S_3 = s_3 + 0$.
Thus after spontaneous symmetry breaking the scalar potential $V$ 
takes the form
\bea
V&=& \frac{\mu_H^2} {2} (v + h)^2 + \frac{\lambda_H} {4} (v + h)^4 +
\frac{\mu_{S_2}^2} {2} s_2^2 + \nonumber \\
&& \frac{\lambda_{S_2}} {4} s_2^4 +
\frac{\mu_{S_3}} {2} s_3^2 + \frac{\lambda_{S_3}} {4} s_3^4 + \nonumber\\
&& \frac{\lambda_{HS_2}} {2} (v + h)^2 s_2^2 +
\frac{\lambda_{HS_3}}{2} (v + h)^2 s_3^2 +
\lambda_{S_2 S_3} s_2^2 s_3^2 \,\, .
\label{eq7}
\eea
Now by using the minimisation condition
\be
\left (\frac{\partial V}{\partial h}\right),\,
\left(\frac{\partial V}{\partial s_2}\right),\,
\left ( \frac {\partial V} {\partial s_3} \right)\Bigg \vert_
{h = 0,\, s_2=0,\,s_3=0}
=0 \,\, ,
\ee
we obtain the condition 
\bea
\mu_H^2 + \lambda_H v^2 = 0\,\, .
\label{min}
\eea
By evaluating
$\frac {\partial^2 V} {\partial h^2}$,
$\frac {\partial^2 V} {\partial s_2^2}$,
$\frac {\partial^2 V} {\partial s_3^2}$,
$\frac {\partial^2 V} { { \partial h}{\partial s_2}}$,
$\frac {\partial^2 V} { { \partial h}{\partial s_3}}$,
$\frac {\partial^2 V} { { \partial s_3}{\partial s_2}}$
at $h=s_2=s_3=0$,  one can now 
construct the mass  matrix in the basis $h-s_2-s_3$ as
\bea
\mathcal{M}^2_{\rm scalar} &=& \left(\begin{array}{ccc}
2\lambda_H v^2 & 0 & 0 \\
0 & \mu_{S_2}^2 + \lambda_{H S_2}v^2 & 0  \\
0 & 0 & \mu_{S_3}^2 + \lambda_{H S_3}v^2
\end{array}
\right).
\label{mat}
\eea
It may be noted here that the mass matrix is diagonal as there is no mixing
between $h$, $s_2$ and $s_3$.

\section{Constraints}
In this section we discuss various bounds and constraints on
the model parameters of the model from both theoretical considerations and
experimental observations. These are furnished in the following.
\vskip 2mm
\noindent $\bullet$ \textbf {Vacuum Stability}:
In our work we consider an extended model with two additional scalar fields. 
For the stability of the vacuum, the scalar potential has to be bounded from 
below in the limit of large field values along all possible directions of the 
field space. In this large
limit the quartic terms of the scalar potential dominate over the mass and the
cubic terms. The quartic part ($V_4$) of the scalar potential $V$
(Eq. (\ref{potential})) is given as
\bea
V_4 &=& \lambda_{H}\,(H^\dagger H)^2 + \frac {\lambda_{S_2}}{4} S_2^4
+ \frac{\lambda_{S_3}}{4} S_3^4 + {\lambda_{H S_2}} H^\dagger H\, S_2^2\nonumber\\
&&+ {\lambda_{HS_3}} H^\dagger H\,S_3^2 + \lambda_{S_2 S_3} S_2^2\,S_3^2 \,.
\label{potential1}
\eea
Bounds on the couplings from the vacuum stability condition are \cite{kannike}
\bea
\lambda_H,\, \, \lambda_{S_2},\,\, \lambda_{S_3} &>& 0 \nonumber \\
\lambda_{H S_2} + \sqrt {\lambda_H \lambda_{S_2}} & > & 0 \nonumber \\
\lambda_{HS_3} + \sqrt {\lambda_H \lambda_{S_3}} & > & 0 \nonumber \\
2\lambda_{S_2 S_3} + \sqrt {\lambda_{S_2} \lambda_{S_3}} & > & 0
\eea
and
\bea
&&\sqrt{2(\lambda_{HS_2} + \sqrt {\lambda_H \lambda_{S_2}})
(\lambda_{HS_3} + \sqrt {\lambda_H \lambda_{S_3}})
(2\lambda_{S_2 S_3} + \sqrt {\lambda_{S_2} \lambda_{S_3}})}\nonumber \\
&&+\sqrt {\lambda_H \lambda_{S_2} \lambda_{S_3}}
+\lambda_{HS_2} \sqrt {\lambda_{S_3}} + \lambda_{HS_3} \sqrt {\lambda_{S_2}} +
2\lambda_{S_2 S_3} \sqrt{\lambda_H}> 0 \,\, .
\label{vac}
\eea
\vskip 2mm
\noindent $\bullet$ \textbf {Perturbativity}:
In order to obey the perturbative limit, the quartic couplings of the 
scalar potential in our model should be constrained as 
\cite{kamakshya}-\cite{lee}
\bea
\lambda_H,\,\, \lambda_{HS_2},\,\, \lambda_{HS_3} \leq 4\pi,\nonumber\\
\lambda_{S_2},\,\, \lambda_{S_3},\,\, \lambda_{S_2 S_3} \leq \displaystyle\frac {2\pi} {3}.
\label{perturbitivity}
\eea
\vskip 2mm
\noindent $\bullet$ \textbf {Relic Density}:
The total relic density of the dark matter components must satisfy
PLANCK observational results for dark matter relic densities.
\be
0.1172 \leq \Omega_{\rm DM} {\tilde {h}}^2 \leq 0.1226\,\, ,
\ee
where $\Omega_{\rm DM}$ is the dark matter relic density normalised to 
the critical density of the Universe and ${\tilde {h}}$ is the Hubble
parameter in units of 100 Km s$^{-1}$ Mpc$^{-1}$.
\vskip 2mm
\noindent $\bullet$ \textbf {Collider Physics Bounds}:
ATLAS and CMS had observed independently the excess in $\gamma \gamma$
channel from which they had confirmed the existence of a Higgs
like scalar with mass $\sim$ 125.5 GeV \cite{Aad,Chat}. 
The signal strength of Higgs like boson is defined as
\bea
R &=& \displaystyle\frac {\sigma (pp \rightarrow h)} {\sigma^{\rm {SM}}
(pp \rightarrow h)} \displaystyle\frac{ {\rm {Br}} (h \rightarrow xx)}
{ {\rm {Br}}^{\rm {SM}} (h \rightarrow xx)}\,\,\, ,
\label{signalstrength}
\eea
where $\sigma (pp \rightarrow h)$ and ${\rm Br} (h \rightarrow xx)$ denote the
production cross-section and the decay branching ratio of Higgs like 
particle decaying into SM particles ($x$) respectively while 
$\sigma^{\rm SM} (pp \rightarrow h)$ and 
${\rm Br}^{\rm SM} (h \rightarrow xx)$ respectively are those for SM Higgs.
The braching ratio of Higgs like boson and SM Higgs boson 
can be expressed respectively as
${\rm Br} (h \rightarrow xx) = \displaystyle\frac {\Gamma (h \rightarrow xx)}
{\Gamma}$ and ${\rm Br}^{\rm SM} (h \rightarrow xx) = \displaystyle\frac 
{\Gamma^{\rm SM} (h \rightarrow xx)} {\Gamma^{\rm SM}}$, where
$\Gamma (h \rightarrow xx)$ and $\Gamma^{\rm SM} (h \rightarrow xx)$  are the
decay width of Higgs like boson and SM Higgs boson. The quantities $\Gamma$ and 
$\Gamma^{\rm SM}$ represent the total decay widths of Higgs like particle  
and SM Higgs boson respectively. Using these expressions for branching 
ratio in Eq. (\ref{signalstrength}) one obtains
\bea
R = \displaystyle\frac {\sigma (pp \rightarrow h)} {\sigma^{\rm SM} (pp \rightarrow h)}
\displaystyle\frac {\Gamma (h \rightarrow xx)} {\Gamma}
\displaystyle\frac {\Gamma^{\rm SM}} {\Gamma^{\rm SM} (h \rightarrow xx)}\,\, .
\label{sigstren}
\eea

As there is no mixing between the scalars ($h$, $s_2$ and $s_3$) we have 
$\sigma (pp \rightarrow h) \equiv \sigma^{\rm SM} 
(pp \rightarrow h)$ and similarly
$\Gamma (h \rightarrow xx) \equiv \Gamma^{\rm SM} (h \rightarrow xx)$.  
Thus Eq. (\ref{sigstren}) takes the form
\bea
R = \displaystyle\frac {\Gamma^{\rm SM}} {\Gamma}\,\, .
\label{sig}
\eea
In the above expressions the total decay width of Higgs like boson 
can be wriiten as
$\Gamma =  \Gamma^{\rm SM} + \Gamma^{\rm inv}$. The invisible decay
width of Higgs like boson to dark matter particles $\Gamma^{\rm inv}$ is given as
\bea
\Gamma^{\rm inv} = \Gamma_{h \rightarrow s_2s_2} + \Gamma_{h \rightarrow s_3s_3}\,\, .
\label{invdecay}
\eea
The decay width $\Gamma_{h \rightarrow s_is_i}$ ($i =2,3$) can be expressed as
\bea
\Gamma_{h \rightarrow s_is_i} = \displaystyle\frac {\lambda_{h s_i s_i}} {8 \pi m_h}
\sqrt{1 - \displaystyle\frac {4m^2_{s_i}} {m_h^2}}\,\, .
\label{decayh}
\eea
The invisible branching ratio for such invisible decay is then given as
\bea
{\rm Br}^{\rm inv} = \displaystyle\frac {\Gamma^{\rm inv} (h \rightarrow s_is_i)}
{\Gamma_h}, i =2,3\,\, .
\label{branchinv}
\eea
We have checked that due to the small values of the couplings in our model this
branching ratio (Eq. (\ref{branchinv})) for the invisible decay of Higgs like boson 
has to be small. 
To this end, we impose the condition 
${\rm Br}^{\rm inv} < 0.2$ \cite{belanger} and that the Higgs like boson  
signal strength must satisfy the limit $R \ge 0.8$ \cite{atlas}. 

\section{Relic Density Calculations for Two Component Scalar 
FIMP Dark Matter}
The evolution of the number density of DM particle with time is governed by 
the Boltzmann equation. In this section we compute the number densities for 
both the DM candidates $s_2$ and $s_3$ in our model, at the present epoch 
(temperature $T_0$ $\sim$ $10^{-13}$ GeV). For the case of a two component 
dark matter, the relic density is obtained  by solving self consistently, two 
coupled Boltzmann equations which, for the present scenario, are given by
\bea
\displaystyle\frac {dn_{s_2}} {dt} + 3{\tilde {H}} n_{s_2} &=& -{\langle \Gamma_{h \rightarrow s_2s_2} \rangle}(n_{s_2} - n_{s_2}^{\rm eq})
-{\langle \sigma v \rangle}_{s_2 s_2 \rightarrow x \bar{x}} (n_{s_2}^2 
 - (n_{s_2}^{\rm eq})^2) \nonumber\\
&& - {\langle \sigma v \rangle}_{s_2 s_2 \rightarrow s_3 s_3} \left(n_{s_2}^2 - 
\displaystyle\frac{(n_{s_2}^{\rm eq})^2} {(n_{s_3}^{\rm eq})^2} n_{s_3}^2 
\right)\,\, ,
\label{eq1}
\eea
\bea
\displaystyle\frac {dn_{s_3}} {dt} + 3{\tilde {H}} n_{s_3} 
&=& -{\langle \Gamma_{h \rightarrow s_3s_3} \rangle}(n_{s_3} - n_{s_3}^{\rm eq})-{\langle \sigma v \rangle}_{s_3 s_3 \rightarrow x \bar{x}} 
(n_{s_3}^2 - (n_{s_3}^{\rm eq})^2) \nonumber \\
&& + {\langle \sigma v \rangle}_{s_2 s_2 \rightarrow s_3 s_3}\left (n_{s_2}^2 - 
\displaystyle\frac{(n_{s_2}^{\rm eq})^2} {(n_{s_3}^{\rm eq})^2} n_{s_3}^2 
\right)\,\, .
\label{eq2}
\eea
In the above, $n_{s_i}$ and $n^{\rm eq}_{s_i}$ ($i=2,3$) are the number densities
(that evolve with time $t$) and equilibrium number densities respectively 
for the scalars $s_2$ and $s_3$,
$\langle \sigma v \rangle_{s_i s_i} \rightarrow ab$ denotes the average 
annihilation cross-sections for the two scalars $s_i, i=2,3$ ($a$, $b$ are 
the annihilation products) and ${\tilde {H}}$ is the Hubble parameter. 

This is to mention that the Boltzmann equations (Eqs. (\ref{eq1}, \ref{eq2}) )
should also in principle include terms due to $4 \rightarrow 2$ or $3 \rightarrow2$ interactions 
of the dark matter self annihilations. The annihilation cross-sections for such processes 
such as $s_2s_2s_2s_2 \rightarrow s_2s_2, s_2s_2s_3s_3 \rightarrow s_2s_3, s_2s_2s_2 \rightarrow s_2s_2, 
s_2s_2s_3 \rightarrow s_2s_3$ could be significant if the couplings are large. For our cases we consider FIMP dark matter masses in three ranges namely keV, MeV and GeV while for GeV range such contributions are ruled out since for a significant contribution, the coupling is to be large enough that may violate perturbative limit \cite{matti}. In case of keV range we have checked (also by Ref. \cite{madhu}) that $4 \rightarrow 2$ interaction is insignificant due to smalleness of corresponding self coupling while for MeV range FIMP these could be significant. We have checked that for the chosen mass and the values of the couplings (obtained from theoretical constraints) the contribution is negligibly small even for MeV mass range FIMPs. From Fig. 3 of Ref. \cite{simp}, we see that for the present work the contribution for MeV mass range falls in the semirelativistic region of the plot. Hence we did not consider these terms in the Boltzmann equations.

\begin{figure}
\begin{center}
\includegraphics[height=12 cm, width=17 cm, angle=0]{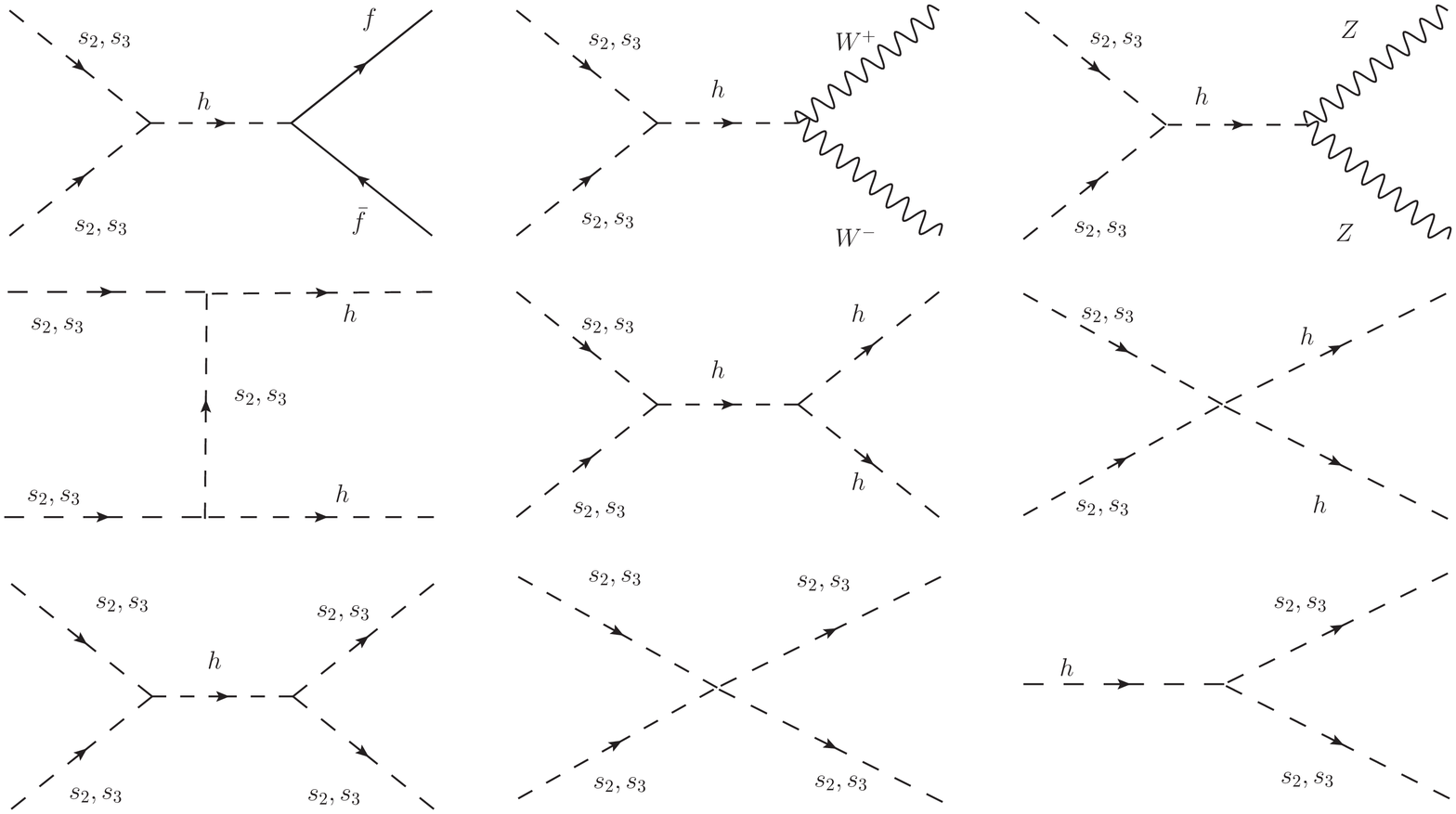}
   \caption{{ Feynman diagrams for both the scalar dark matter candidates $s_2$ and $s_3$.}}
   \label{a}
\end{center}
\end{figure}

Defining a dimensionless quantity namely the comoving number 
density expressed in terms of the ratio $Y_i = n_{s_i}/S$ ($i=2,3$) of
the number density ($n_{s_i}$) and the total entropy density ($S$)
and defining $z = m_h/T$, $T$ being the photon temperature,  Eqs. (\ref{eq1},
\ref{eq2}) can be rewritten in terms of the variation of $Y_i (i=2,3)$ with $z$ as
\bea
\displaystyle\frac {dY_{s_2}} {dz} &=& 
-\displaystyle\frac {2zm_{\rm Pl}} {1.66 m_h^2} \displaystyle\frac {\sqrt {g_*(T)}} {\sqrt {g_S(T)}} \Bigg(\langle \Gamma_{h 
\rightarrow s_2 s_2} \rangle \left ( Y_{s_2} - Y_h^{\rm eq} \right)\Bigg) 
\nonumber\\
&& - \displaystyle\frac {4\pi^2} {45} \displaystyle\frac {m_{\rm Pl} m_h} 
{1.66}
\displaystyle\frac {\sqrt{g_*(T)}} {z^2} \times \nonumber\\
&&\Bigg ( \sum_{x = W,Z,f,h} 
\langle {\sigma v}_{x \bar{x} \rightarrow s_2 s_2} \rangle 
(Y_{s_2}^2 - ({Y_x^{\rm eq}})^2 ) + \langle {\sigma v}_{s_2 s_2 
\rightarrow s_3 s_3} 
\rangle (Y_{s_2}^2 - \displaystyle\frac {(Y_{s_2}^{\rm eq})^2}
{(Y_{s_3}^{\rm eq})^2} Y_{s_3}^2 )  \nonumber\\
&& - \langle {\sigma v}_{s_3 s_3 
\rightarrow s_2 s_2} \rangle (Y_{s_3}^2 - 
\displaystyle\frac {(Y_{s_3}^{\rm eq})^2}
{(Y_{s_2}^{\rm eq})^2} Y_{s_2}^2)\Bigg ) 
\label{eq3}
\eea
and
\bea
\displaystyle\frac {dY_{s_3}} {dz} &=& -\displaystyle\frac {2zm_{\rm Pl}} 
{1.66 m_h^2} \displaystyle\frac {\sqrt {g_*(T)}} {\sqrt {g_S(T)}} 
\Bigg(\langle \Gamma_{h\rightarrow s_3 s_3} 
\rangle \left ( Y_{s_3} - Y_h^{\rm eq} \right)\Bigg) \nonumber\\
&& - \displaystyle\frac {4\pi^2} {45} \displaystyle\frac {m_{\rm Pl} m_h} 
{1.66} \displaystyle\frac {\sqrt{g_*(T)}} {z^2} \times \nonumber\\
&&\Bigg ( \sum_{x = W,Z,f,h}
\langle {\sigma v}_{x \bar{x} \rightarrow s_3 s_3} 
\rangle (Y_{s_3}^2 - ({Y_x^{\rm eq}})^2 ) -
\langle {\sigma v}_{s_2 s_2 \rightarrow s_3 s_3}
\rangle (Y_{s_2}^2 - \displaystyle\frac {(Y_{s_2}^{\rm eq})^2}
{(Y_{s_3}^{\rm eq})^2} Y_{s_3}^2 )  \nonumber\\
&& + \langle {\sigma v}_{s_3 s_3
\rightarrow s_2 s_2} \rangle (Y_{s_3}^2 -
\displaystyle\frac {(Y_{s_3}^{\rm eq})^2}
{(Y_{s_2}^{\rm eq})^2} Y_{s_2}^2)\Bigg )\,\, .
\label{eq4}
\eea
We have already mentioned that the initial abundance of 
FIMP \cite{McDonald:2001vt,Hall:2009bx} dark matter candidate is 
negligible. Therefore assuming $Y_{s_2} = Y_{s_3} = 0$, 
Eqs. (\ref{eq3}, \ref{eq4}) take the form
\bea
\displaystyle\frac {dY_{s_2}} {dz} &=& -\displaystyle\frac {2zm_{\rm Pl}} 
{1.66 m_h^2} \displaystyle\frac {\sqrt {g_*(T)}} {\sqrt {g_S(T)}} 
\Bigg(\langle \Gamma_{h\rightarrow s_2 s_2} 
\rangle \left (  - Y_h^{\rm eq} \right)\Bigg) \nonumber\\
&& - \displaystyle\frac {4\pi^2} {45} \displaystyle\frac {m_{\rm Pl} m_h} 
{1.66} \displaystyle\frac {\sqrt{g_*(T)}} {z^2} \times \nonumber\\
&&\Bigg ( \sum_{x = W,Z,f,h}
\langle {\sigma v}_{x \bar{x} \rightarrow s_2 s_2} 
\rangle (- {Y_x^{\rm eq}})^2  \Bigg)
\label{eq5}
\eea
and
\bea
\displaystyle\frac {dY_{s_3}} {dz} &=& -\displaystyle\frac {2zm_{\rm Pl}} {1.66 m_h^2} \displaystyle\frac {\sqrt {g_*(T)}} {\sqrt {g_S(T)}} \Bigg(\langle \Gamma_{h
\rightarrow s_3 s_3} \rangle \left (  - Y_h^{\rm eq} \right)\Bigg) \nonumber\\
&& - \displaystyle\frac {4\pi^2} {45} \displaystyle\frac {m_{\rm Pl} m_h} {1.66}
\displaystyle\frac {\sqrt{g_*(T)}} {z^2} \times \nonumber\\
&&\Bigg ( \sum_{x = W,Z,f,h}
\langle {\sigma v}_{x \bar{x} \rightarrow s_3 s_3} \rangle (- {Y_x^{\rm eq}})^2   \Bigg)\,\, .
\label{eq6}
\eea
In the above, $m_{\rm Pl}$ is the PLANCK mass, $m_{\rm Pl}$ = 1.22 $\times 
10^{22}$ GeV and the term $g_*$ is defined as \cite{Gondolo:1990dk}
\bea
\sqrt {g_*(T)} = \displaystyle\frac {g_S(T)} {\sqrt{g_{\rho}{T}}}\Bigg ( 1 + \displaystyle\frac {1} {3} \displaystyle\frac {d lnh_{\rm eff}(T)} 
{d lnT}\Bigg)\,\, ,
\label{eq7}
\eea
where two effective degrees of freedom $g_{\rm eff}(T)$ and $h_{\rm eff}(T)$ 
are related to the energy and entropy densities of the Universe through the 
following relations,
\bea
S = g_S(T) \displaystyle\frac {2\pi^2} {45} T^3\, , &&
\rho = g_{\rho}(T) \displaystyle\frac {\pi^2} {30} T^4\,\, .
\label{eq8}
\eea
Also, the thermally averaged decay widths and annihilation cross-sections 
for various processes are given by,
\bea
\langle \Gamma_{h\rightarrow s_i s_i} \rangle&=& \Gamma_{h\rightarrow
s_i s_i}\frac{K_1(z)}{K_2(z)}\,\, ,\nonumber\\ 
\langle \sigma {\rm v} \rangle_{x \bar{x}\rightarrow
s_i s_i}&=&\frac{1}{8m_x^4\,T\,K_2^2(M_x/T)}
\int_{4m_x^2}^\infty \sigma_{x \bar{x}\rightarrow s_i s_i}(s-4M_x^2)\sqrt{s}
K_1(\displaystyle\frac{\sqrt{s}} {T} )ds\,\, .
\label{eq9}
\eea
In Eq. (\ref{eq9}) $i = 2,3 , x = W^{\pm},Z,f,h,s_2,s_3$, 
$K_1$ and $K_2$ are the modified Bessel functions of order 1 and 2, $s$ defines the centre of momentum energy. The decay widths 
$\Gamma_{h \rightarrow S_iS_i}$ and annihilation
cross-sections $\sigma_{x \bar{x} \rightarrow s_is_i}$ ($i = 2,3$) for
different processes considered to calcuate the coupled Boltzmann
equations (Eqs. (\ref{eq5}, \ref{eq6})) are given below
\bea
\Gamma_{h \rightarrow s_is_i} &=& \displaystyle\frac {g^2_{hs_is_i}} {8\pi m_h}
\sqrt{1 - \displaystyle\frac {4 m^2_{s_i}} {m^2_h}}\,\, , \\
\sigma_{hh \rightarrow s_is_i} &=& \displaystyle\frac {1} {2\pi s}
\sqrt{\displaystyle\frac {s - 4m^2_{s_i}} {s - 4m^2_h}} \left \{ g^2_{hhs_is_i}
+ \displaystyle\frac {9\,\, g^2_{hhh} g^2_{hs_is_i}} {[ (s - m^2_h)^2 +
(\Gamma_h m_h)^2] }\right. \nonumber\\
&&\left. - \displaystyle \frac {6\,\, g_{hhs_is_i} g_{hs_is_i}
g_{hhh} (s - m^2_h)} { [ (s - m^2_h)^2 + (\Gamma_h m_h)^2 ]}
\right\}\,\, , \\
\sigma_{s_2s_2 \rightarrow s_3s_3} &=& \displaystyle\frac {1} {2\pi s}
\sqrt{\displaystyle\frac {s - 4m^2_{s_3}} {s - 4m^2_{s_2}}} \left
\{ g^2_{s_2s_2s_3s_3} 
+ \displaystyle\frac { g^2_{s_2s_2h} g^2_{hs_3s_3}} {[ (s - m^2_h)^2 +
(\Gamma_h m_h)^2] } \right . \nonumber\\
&& \left . - \displaystyle \frac {2\,\, g_{s_2s_2s_3s_3} g_{s_2s_2h}
g_{hs_3s_3} (s - m^2_h)} { [ (s - m^2_h)^2 + (\Gamma_h m_h)^2 ]}
\right \}\,\, , \\
\sigma_{s_3s_3 \rightarrow s_2s_2} &=& \displaystyle\frac {1} {2\pi s}
\sqrt{\displaystyle\frac {s - 4m^2_{s_2}} {s - 4m^2_{s_3}}} \left
\{ g^2_{s_2s_2s_3s_3}
+ \displaystyle\frac { g^2_{s_3s_3h} g^2_{hs_2s_2}} {[ (s - m^2_h)^2 +
(\Gamma_h m_h)^2] } \right . \nonumber\\
&& \left . - \displaystyle \frac {2 g_{s_3s_3s_2s_2} g_{s_3s_3h}
g_{hs_2s_2} (s - m^2_h)} { [ (s - m^2_h)^2 + (\Gamma_h m_h)^2 ]}
\right\}\,\, , \\
\sigma_{WW \rightarrow s_is_i} &=& \displaystyle\frac {g^2_{WWh} g^2_{hs_is_i}}
{72 \pi s} \sqrt{\displaystyle\frac {s - 4m^2_{s_i}} {s - 4m^2_{W}}}
\displaystyle\frac {\left ( 3 - \displaystyle\frac {s} {m^2_W} +
\displaystyle\frac {s^2} {4m^2_W} \right ) } {(s - m^2_h)^2 +
(\Gamma_h m_h)^2}\,\, , \\
\sigma_{ZZ \rightarrow s_is_i} &=& \displaystyle\frac {g^2_{ZZh} g^2_{hs_is_i}}
{18 \pi s} \sqrt{\displaystyle\frac {s - 4m^2_{s_i}} {s - 4m^2_{Z}}}
\displaystyle\frac {\left ( 3 - \displaystyle\frac {s} {m^2_Z} +
\displaystyle\frac {s^2} {4m^2_Z} \right ) } {(s - m^2_h)^2 +
(\Gamma_h m_h)^2}\,\, , \\
\sigma_{f \bar{f} \rightarrow s_is_i} &=& \displaystyle\frac {N_c g^2_{ffh}
g^2_{hs_is_i}} {16 \pi s} \displaystyle\frac {\sqrt{ (s - 4m^2_{s_i})
(s - 4m^2_f)}} {(s - m^2_h)^2 + (\Gamma_h m_h)^2}\,\, .
\label{decayanni}
\eea
In the above equations the couplings of the vertices are defined as $g_{abc}$
and $g_{abcd}$, where $a, b, c, d $ are the fields. 
The masses of $W$ and $Z$ bosons and the fermions ($f$) are denoted as 
$m_W$, $m_Z$ and $m_f$ respectively. $N_c$ in Eq. (\ref{decayanni}) denotes the color quantum number. Detailed expressions for all the couplings required given in Eqs. (33 - 39) are enlisted in the Appendix. The Feynman diagrams corresponding to all the possible channels for the two distinct scalar components $s_2$ and $s_3$ are shown in Fig. 1.

The relic densities of each of the components $s_2$ and $s_3$ of the dark 
matter are finally obtained in terms 
of their respective masses and comoving number densities at the present 
epoch, as \cite{edsjo,biswas}
\bea
\Omega_{i}{\tilde {h}}^2= 2.755 \times 10^8 
\left(\frac{m_i}{\rm GeV}\right)Y_i(T_0),
\hskip 10pt i= s_2,s_3\,\, .
\label{relic}
\eea
Solving numerically the two coupled Boltzmann equations Eqs. 
(\ref{eq3} - \ref{eq6}) alongwith Eqs. (33 - 39)
we compute the comoving number densities $Y_i(T_0)$ for both components of 
FIMP dark matter. The total relic density $\Omega_{\rm {tot}}$ is then 
obtained by adding the relic densities of each of the components 
$s_2$ and $s_3$ as follows
\be
\Omega_{\rm tot}{\tilde {h}}^2=\Omega_{s_2}{\tilde {h}}^2+
\Omega_{s_3}{\tilde {h}}^2\,\, .
\label{relic2}
\ee
The total relic density $\Omega_{\rm tot} {\tilde {h}}^2$ should 
satisfy the PLANCK measurement
\be
\hskip 10 pt 0.1172\le \Omega_{\rm DM}{\tilde {h}}^2 \le 0.1226\,\, .
\ee

As mentioned earlier, in our present dark matter model we have considered two distinct scalar dark matter particles in the FIMP scenario. In Fig. 2 we furnish representative plots showing the evolutions of relic densities for each of the components as well as the total relic densities of two component scalar dark matter for each of the chosen mass regimes namely GeV (Fig. 2a), MeV (Fig. 2b) and keV (Fig. 2c).

\begin{figure}
\begin{center}
\subfigure[ ]{
\includegraphics[width=6.0 cm, height=6.0 cm, angle=0]{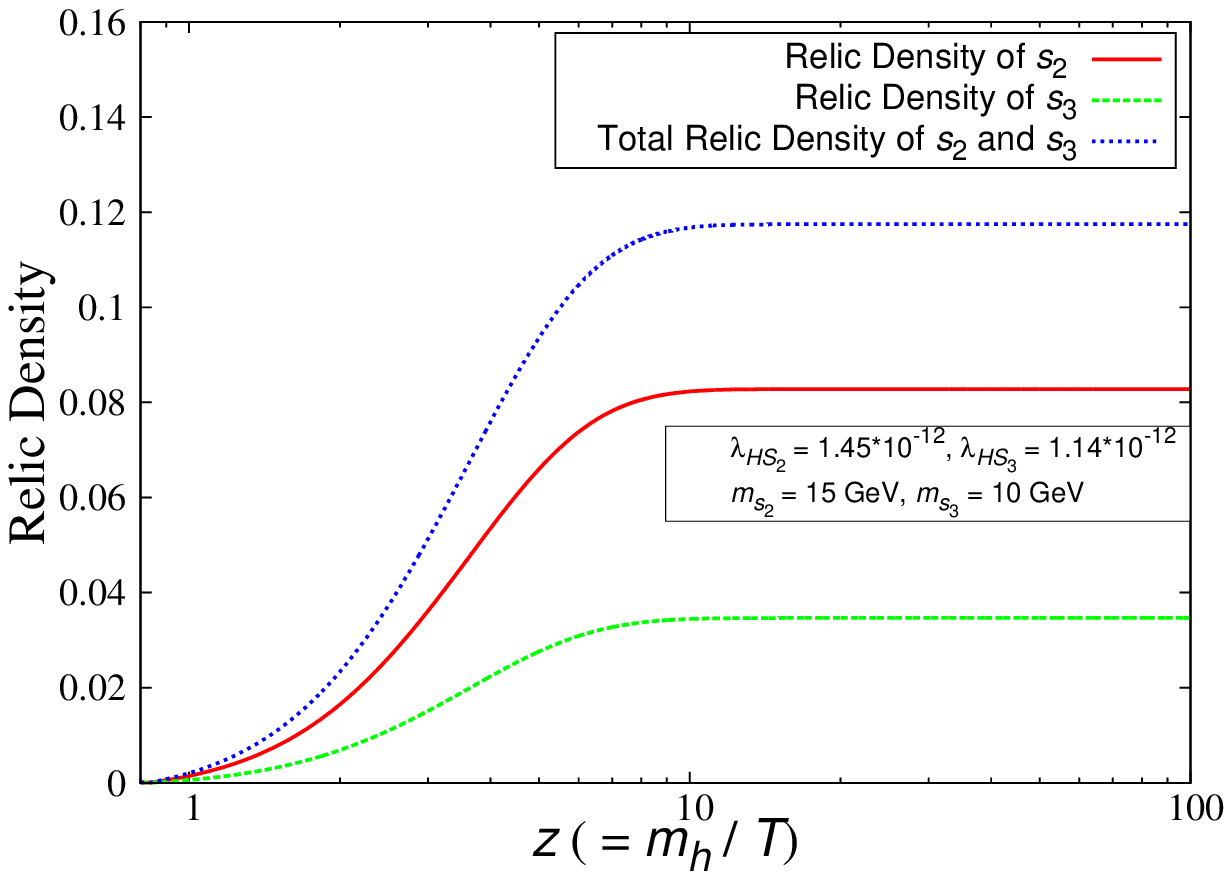}}
\subfigure[ ]{
\includegraphics[width=6.0 cm, height=6.0 cm, angle=0]{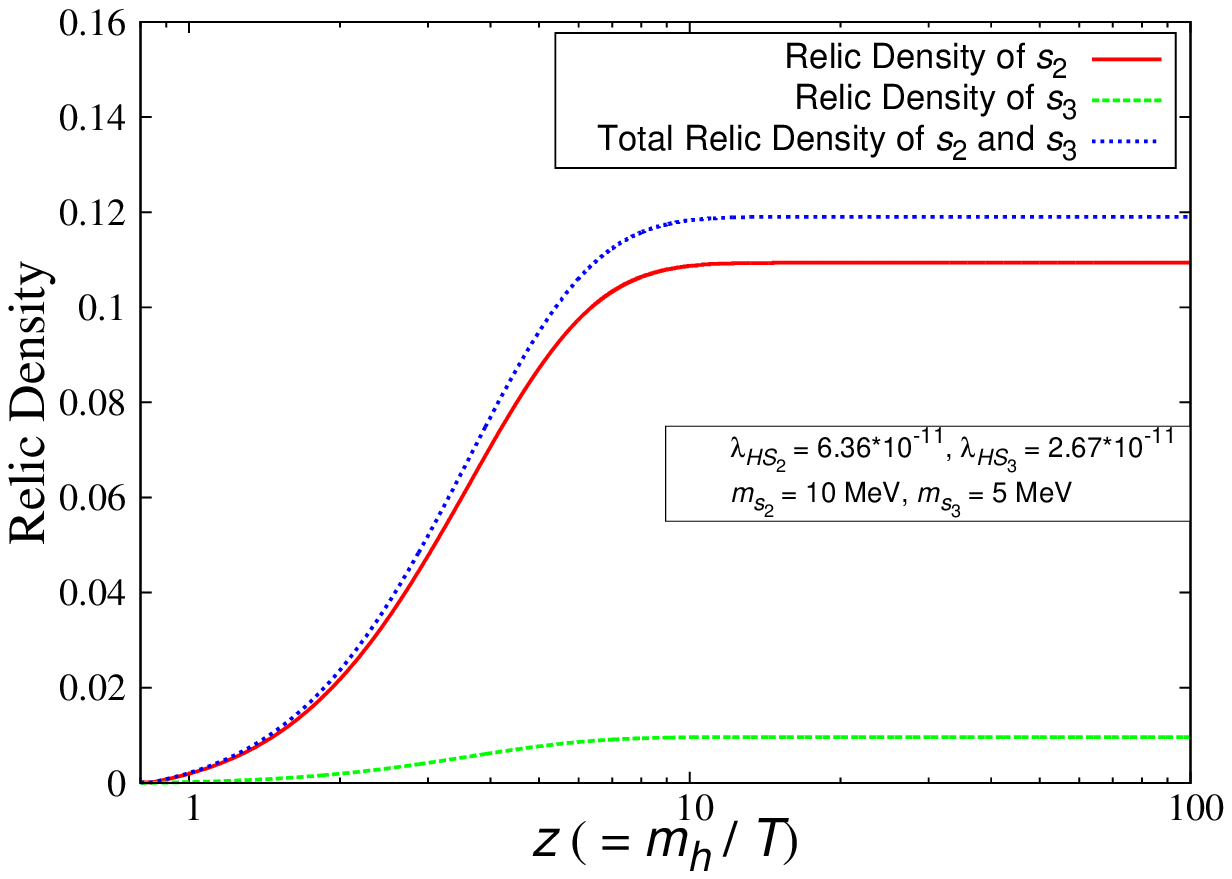}}
\subfigure[]{
\includegraphics[width=6.0 cm, height=6.0 cm, angle=0]{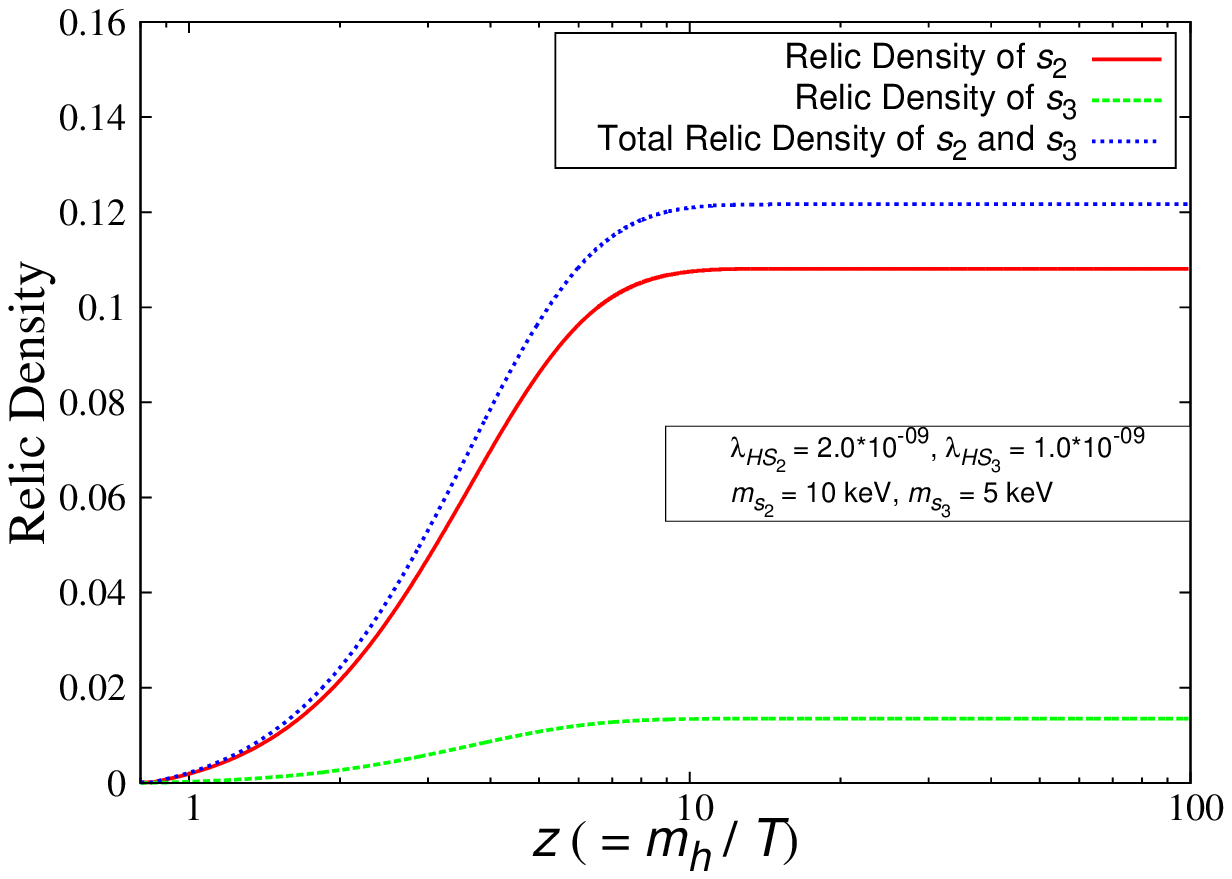}}
  \caption{{Variations of the relic densities of the two single scalar dark matter components $s_2$ and $s_3$ with $z$ for different values of couplings 
($\lambda_{HS_2}, \lambda_{HS_3}$) and masses ($m_{s_2}, m_{s_3}$) in the three mass regions GeV (a), MeV (b) and keV (c) . In each plot (from (a) to (c)) the red solid line and the green dashed line indicate the relic density of the components $s_2$ and $s_3$ respectively  while the blue dotted line represents the same for the total relic density. The PLANCK observational range for DM relic density is $0.1172\le \Omega_{\rm DM}{\tilde {h}}^2 \le 0.1226$. }}
   \label{1}
\end{center}
\end{figure}

\section{Calculations and Results}
We have considered here a two component scalar dark matter model under the 
framework of Feebly Interacting Massive Particle (FIMP) dark matter. 
In this work
this is our purpose to demonstrate that over a wide range of masses (from
GeV to keV) such a two component FIMP scalar dark matter is a viable 
dark matter candidate. Therefore in our analysis 
we have chosen the DM candidates in three mass regimes namely GeV, MeV and keV. 
 To this end we first calculate the relic densities of the FIMP dark 
matter candidates in our proposed model. From Section 3 it should be 
clear that the 
various unknown couplings ($\lambda_{HS_2}, \lambda_{HS_3}, \lambda_{S_2S_3}$ 
etc.) constitute 
the parameters of our model. We first constrain 
those parameters by using various theoretical bounds given in Eqs. (14-16) 
as also the collider bounds described in Eqs. (18-23).
We have chosen a pair of values for two DM components in our two component 
scalar model in each of the three separate mass regimes GeV, MeV and keV. 
The relic densities of each component are first calculated using the 
Eqs. (24-40) by varying the parameter space 
within the constrained range. These are eventually added
up (Eq. (41)) to obtain the total relic density of the present two 
component dark matter model. The expressions for various 
couplings $g_{x_1, x_2,x_3}$ and $g_{x_1,x_2,x_3,x_4}$
(where $x_i$, $i = 1,2,3,4$ represents different particles 
involving annihilation
cross-sections or decay widths) required to compute the relic densities 
by solving 
the Boltzmann equations (Eqs. (26-29)) in terms of the model parameters 
are given in the Appendix. Thus the computed relic densities are then 
compared with the PLANCK 
observational measurements for the same (Eq. (42)). 
Thus the model 
parameter space is 
further constrained by the observed relic denisties for the dark matter. 
We have also checked that the scattering cross-section of each of the 
components of the present model with nucleon is well below the most stringent 
upper bound for the same reported by the LUX dark matter direct detection
experiment \cite{Akerib:2016vxi}. 
In the following 
we describe the calculations for each mass regime considered here. 




\subsection{FIMP at GeV Mass Regime}
In the GeV regime the masses of the two scalar components are chosen 
to be 15 GeV and 10 GeV. The relic densities 
for such two component FIMP dark matter are calculated for each of the 
components by solving the coupled 
Boltzmann equations (Eqs. (26-29)) which are added up to obtain 
the total relic density. The computation 
is performed by varying the model parameters. The range of these 
parameters are so chosen that they satisfy the
theoretical bounds given in Eq. (14-16). This is also verified that 
for the chosen range of the model parameters the collider 
bounds (Sect. 4) are satisfied.
In other words we ensure that within the chosen range of our model 
parameters the signal strength of SM Higgs boson (Eq. (19)) satisfies 
the limit $R \ge 0.8$ and the invisible branching ratio (Eq. (23)) 
satisfies ${\rm Br}^{\rm inv} < 0.2$. 

In Fig. 3 we show the variations of the total relic abundance 
$\Omega_{\rm tot} {\tilde {h}}^2$ (right panel) 
and the relic abundances $\Omega_{s_2, s_3} {\tilde {h}}^2$ 
for each of the components of the present DM model (left panel) 
with $\lambda_{HS_3}$. In Fig. 4 similar variations with coupling
$\lambda_{HS_2}$ are plotted. 
In both the figures the PLANCK observational results for  
$\Omega_{\rm DM} {\tilde {h}}^2$ are shown by two 
parallel lines. It is observed from Figs. 3,4 that 
the relic abundance increases 
with the increase of the parameters 
$\lambda_{HS_3}$ and $\lambda_{HS_2}$. Figs. 3,4 constraints these 
parameters by PLANCK results. In fact from Figs. 3,4 one sees that 
for the chosen fixed FIMP 
component masses of 15 GeV and 10 GeV the upper limits of the 
Higgs-couplings with the scalar components $\lambda_{H S_2}$ and 
$\lambda_{H_{S_3}}$ will be around $10^{-12}$.

\begin{figure}
\begin{center}
\subfigure[ ]{
\includegraphics[width=5.5 cm, height=5.5 cm, angle=-90]{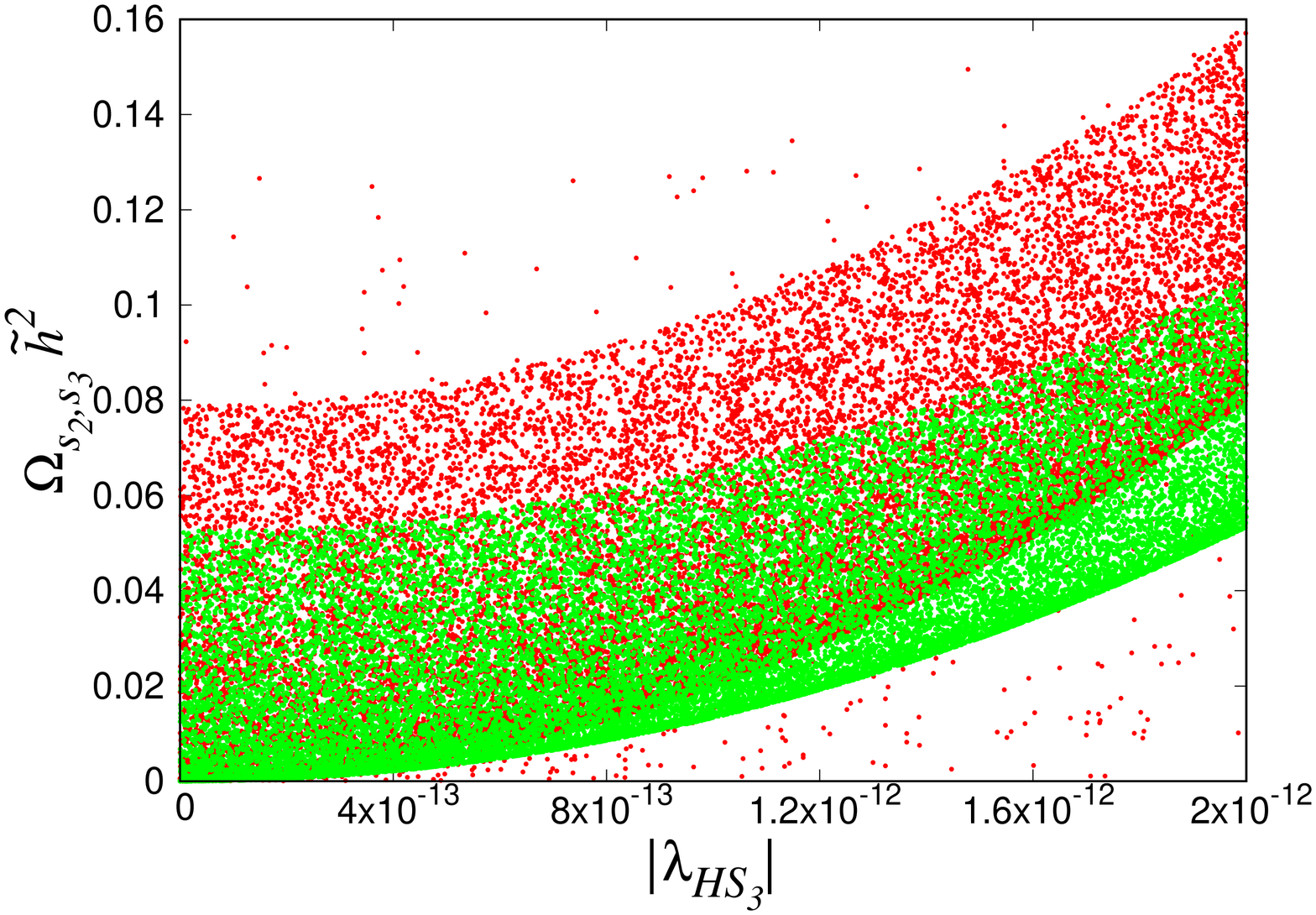}}
\subfigure[ ]{
\includegraphics[width=5.5 cm, height=5.5 cm, angle=-90]{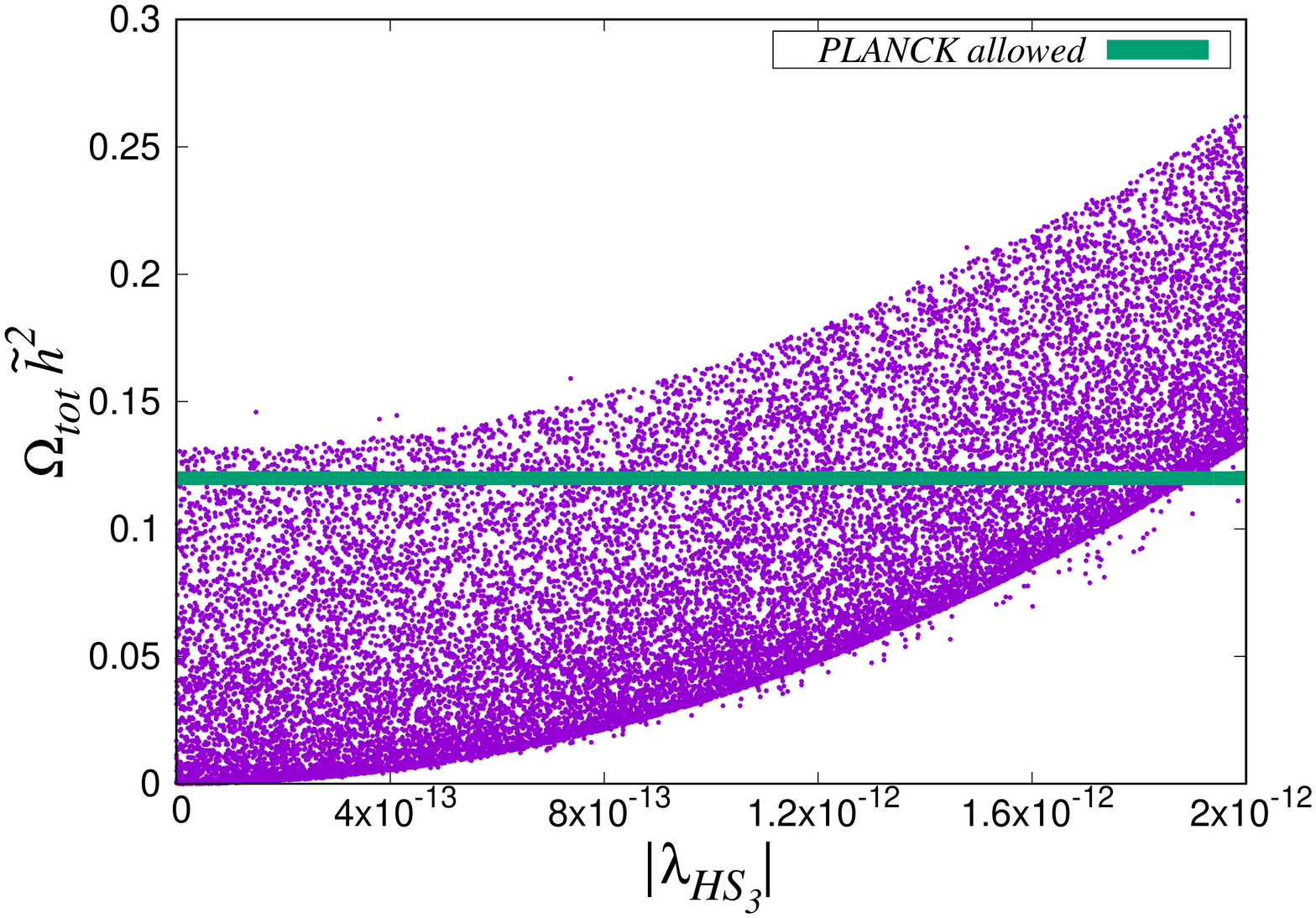}}
  \caption{{The variations of the relic abundances $\Omega_{s_2, s_3} {\tilde {h}}^2$ for each of the two DM components with the coupling $\lambda_{HS_3}$. The red and green regions represent the relic abundaces of $s_2$ and $s_3$ respectively. Right panel shows the variation of $\Omega_{\rm tot} {\tilde {h}}^2$ with $\lambda_{HS_3}$. The PLANCK limit is shown by the thick green line. See text for details. }}
   \label{b}
\end{center}
\end{figure}

\begin{figure}
\begin{center}
\subfigure[ ]{
\includegraphics[width=5.5 cm, height=5.5 cm, angle=-90]{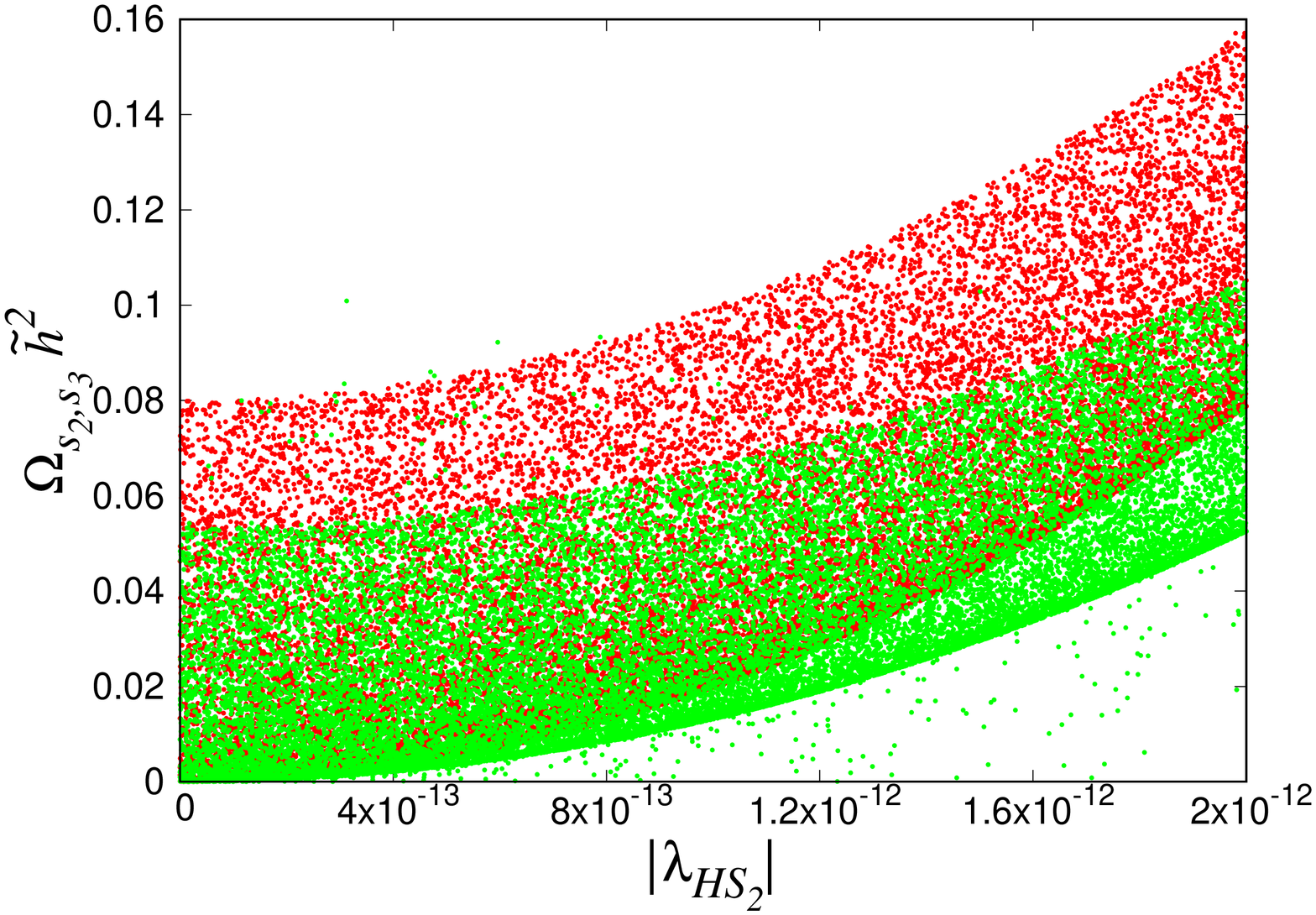}}
\subfigure[ ]{
\includegraphics[width=5.5 cm,height=5.5 cm, angle=-90]{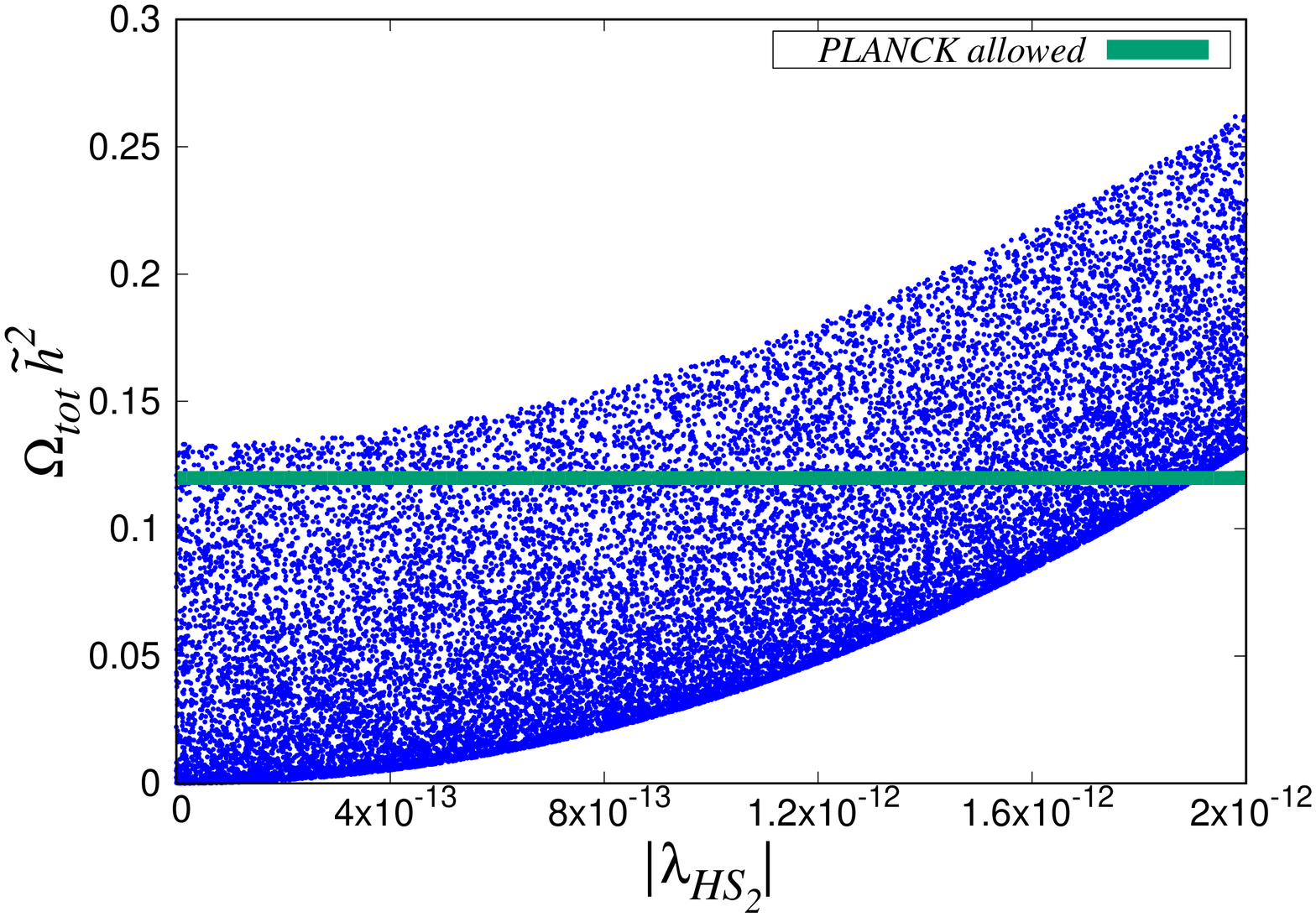}}
   \caption{{Same as Fig. 3 but for $\lambda_{HS_2}$.}}
   \label{b1}
\end{center}
\end{figure}


\begin{figure}
\begin{center}
\subfigure[]{
\includegraphics[width=5.5 cm, height=5.5 cm, angle=-90]{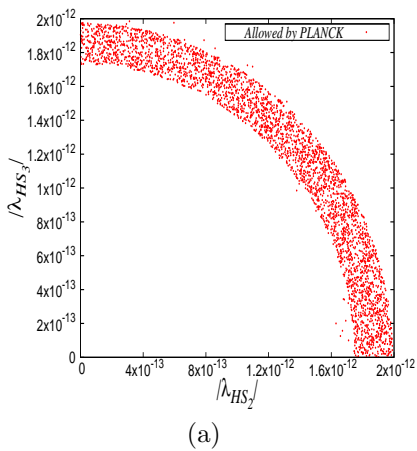}}
  \caption{{The available region constrained by the PLANCK results in $\lambda_{HS_2} - \lambda_{HS_3}$ plane is denoted in this figure.}}
   \label{b3}
\end{center}
\end{figure}

Unlike the WIMP dark matter where
the relic density of dark matter would decrease with the increase of 
the Higgs-couplings with the DM candidates, for the case of FIMP 
DM the relic density increases with the Higgs-couplings instead. 
This is one of the salient features of FIMP dark matter. 
This can also be seen from Figs. 3,4 that the nature of variations of the relic abundances with $\lambda_{HS_i} (i = 2,3)$ 
are parabolic which reflect the fact that $\Omega {\tilde {h}}^2 \sim \lambda_{HS_i}^2 (i = 2,3)$. 


Further, in order to constrain the parameter space by the PLANCK 
observational results we simultaneously vary the 
two parameters $\lambda_{HS_2}, \lambda_{HS_3}$. 
The results are plotted in Fig. 6. 
In Fig. 6 we show the two parameter scan results,
where the region constrained by the PLANCK results are 
shown by the red colour zone in the 
$\lambda_{HS_2} - \lambda_{HS_3}$ plane.

\subsection{FIMP at MeV Mass Regime}
In the MeV regime we choose the masses of the two component dark matter 
to be 10 MeV and 5 MeV.
With these masses and using our formalism of two component FIMP dark matter 
model we constrain the parameter 
space following the procedures similar to what described in 
Section 6.1. In this mass 
region too  the parameter space is finally constrained by calculating 
the relic abundance and then comparing 
them with the PLANCK results. The results are shown in Figs. 6,7. From Fig. 6 and Fig. 7 which show the variations of 
relic abundances for each of the two components as well as the total abundance with the coupling parameters 
$\lambda_{HS_2}, \lambda_{HS_3}$ respectively we obtain an upper limit for $\lambda_{HS_3}$ and $\lambda_{HS_2}$ to be 
of the order of $\sim 8 \times 10^{-11}$. In Fig. 8 we show the parameter space restricted by the PLANCK relic abundance
results and is indicated by the red colour region in the relevant plot.

\subsection{FIMP at keV Mass Regime}
In the keV range we have considered the masses of the two FIMP scalar 
components to be 10 keV and 5 keV
and performed the analysis similar to what described in the cases of 
GeV and MeV ranges for restricting the parameter space. The results 
are shown in Figs. 9,10. 
We find similar nature for variations of the relic abundances with the 
couplings (Figs. 9,10) as also for the constrained 
parameter space (Fig. 11). We find the upper limits for $\lambda_{HS_3}$ 
and $\lambda_{HS_2}$ to be around $2.2 \times 10^{-9}$.

\begin{figure}[h!]
\begin{center}
\subfigure[]{
\includegraphics[width=5.5 cm, height=5.5 cm, angle=-90]{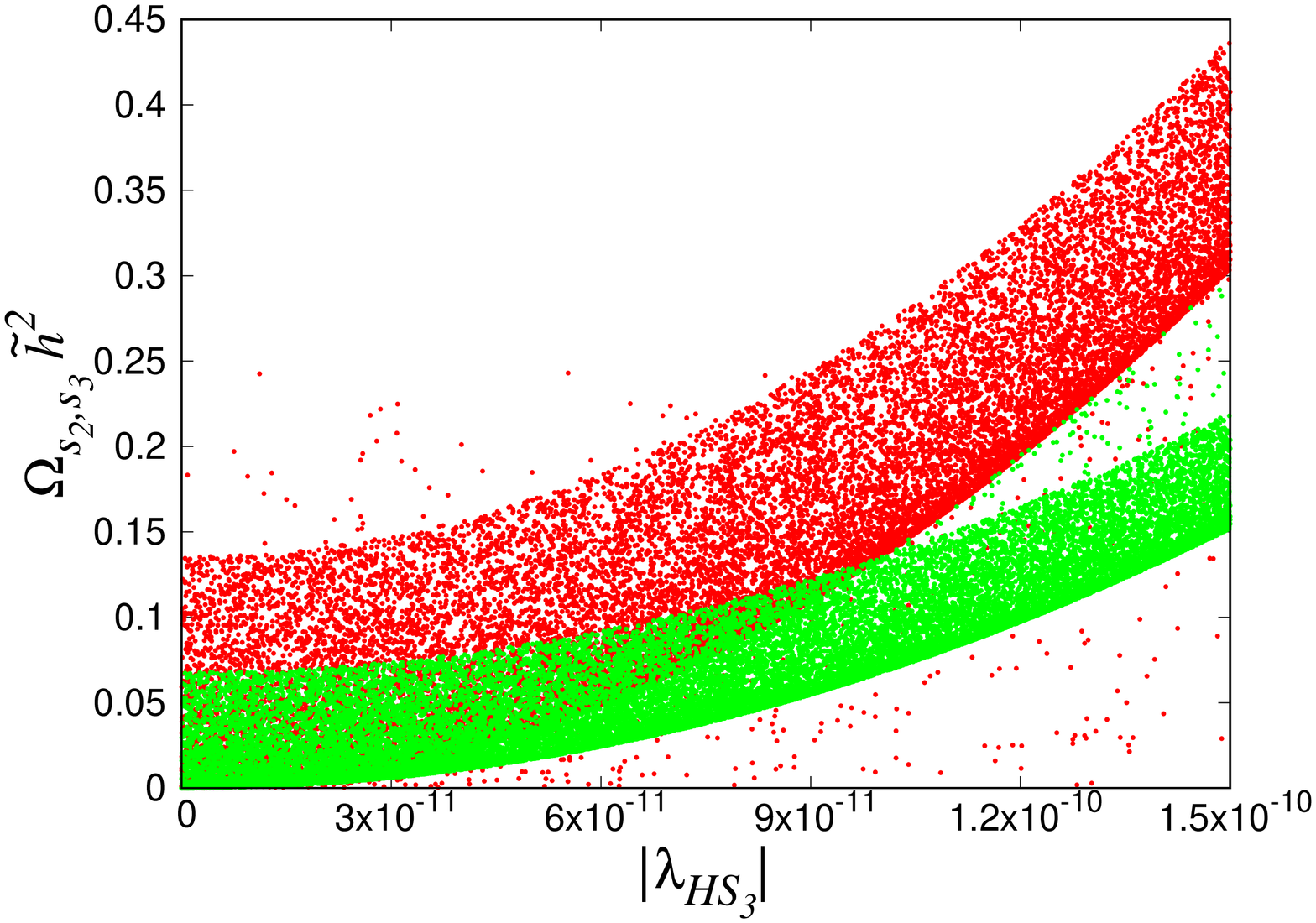}}
\subfigure[]{
\includegraphics[width=5.5 cm, height=5.5 cm, angle=-90]{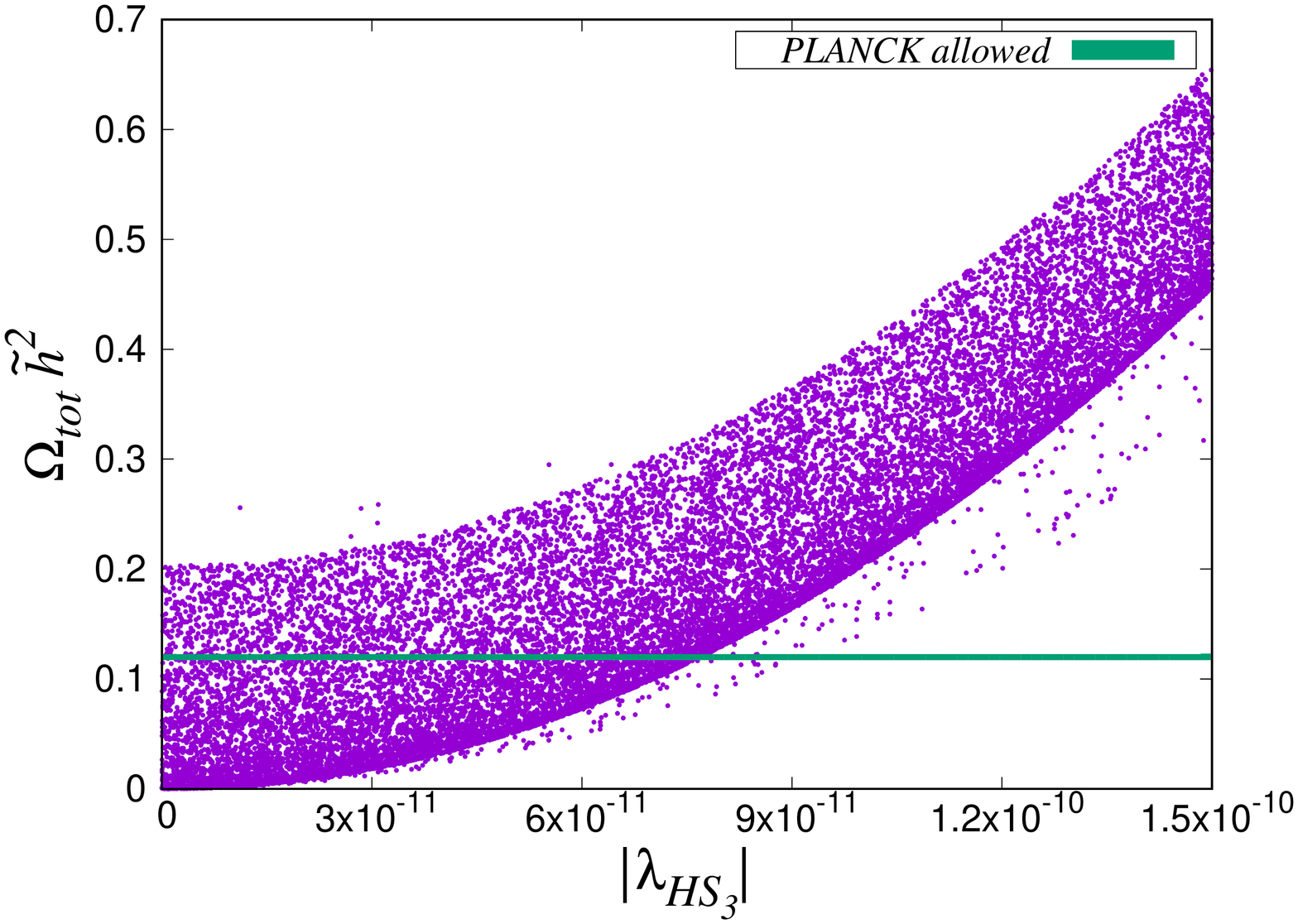}}
   \caption{{Same as Fig. 3 but for MeV mass regime. See text for details.}}
   \label{b5}
\end{center}
\end{figure}

\begin{figure}[h!]
\begin{center}
\subfigure[]{
\includegraphics[width=5.5 cm, height=5.5 cm, angle=-90]{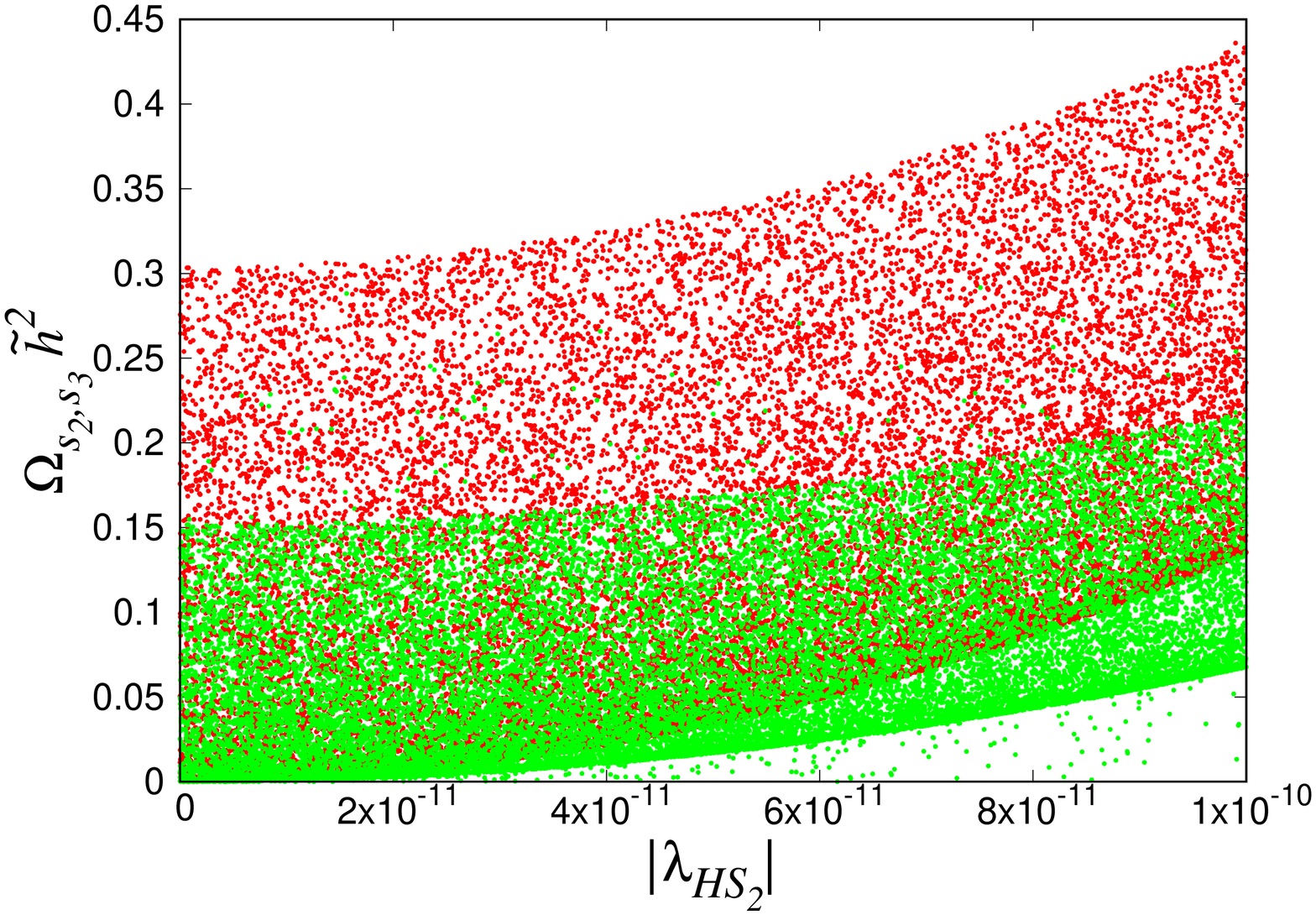}}
\subfigure[]{
\includegraphics[width=5.5 cm, height=5.5 cm, angle=-90]{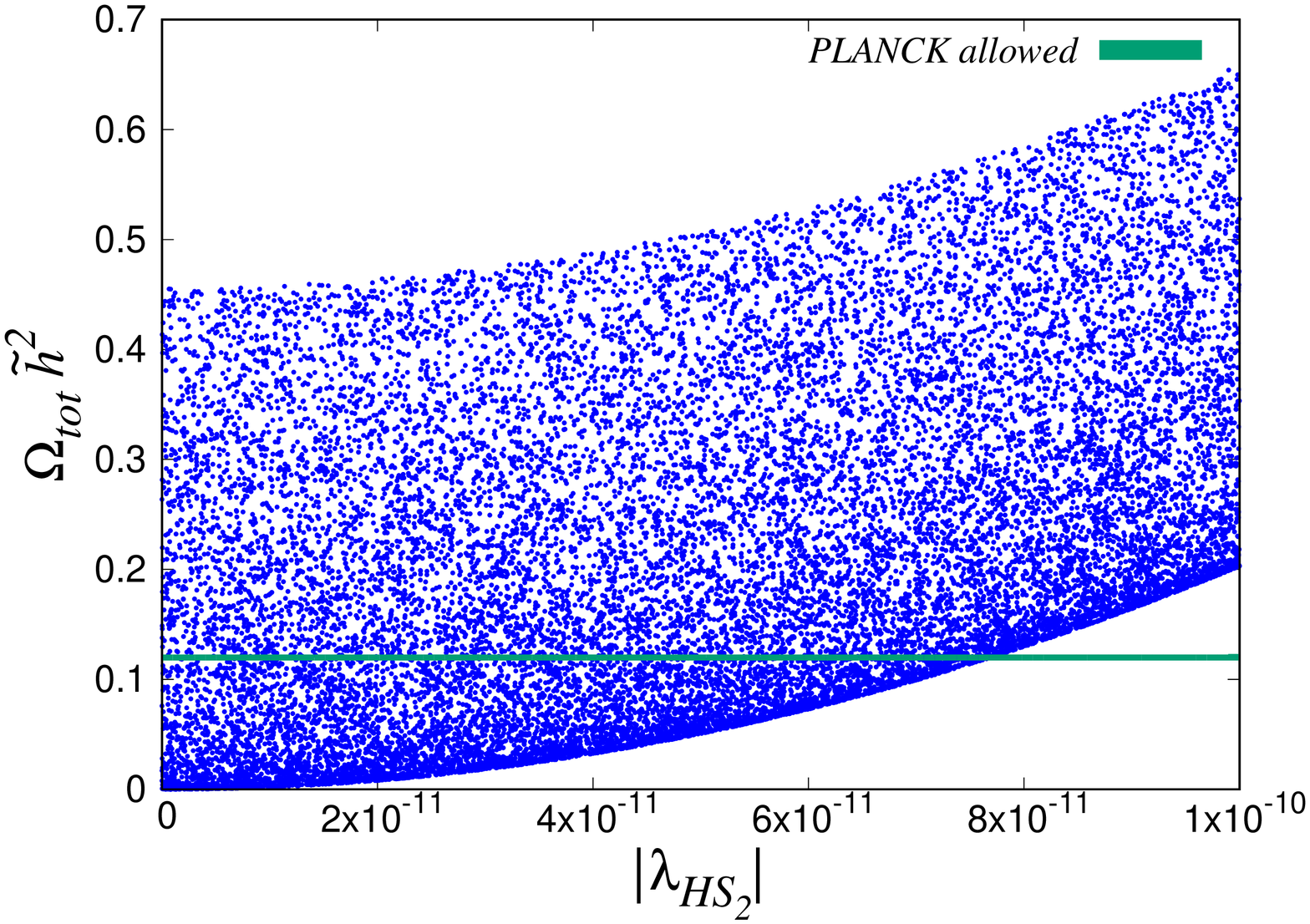}}
   \caption{{Same as Fig. 4 but for MeV mass regime.}}
   \label{b6}
\end{center}
\end{figure}


\begin{figure}[h!]
\begin{center}
\subfigure[]{
\includegraphics[width=5.5 cm, height=5.5 cm, angle=-90]{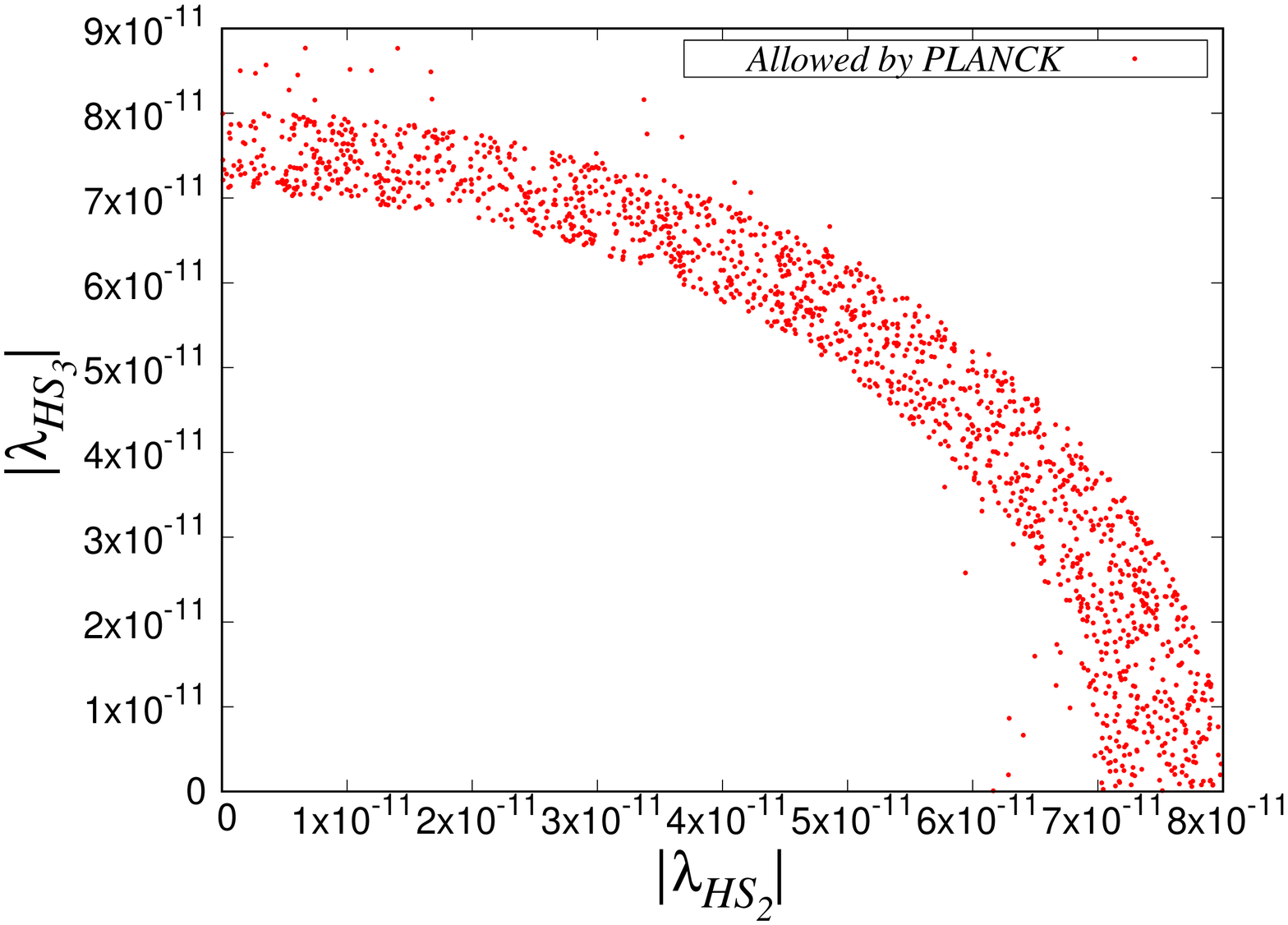}}
   \caption{{Same as Fig. 5 but for MeV mass regime.}}
   \label{b8}
\end{center}
\end{figure}

\begin{figure}[h!]
\begin{center}
\subfigure[]{
\includegraphics[width=5.5 cm, height=5.5 cm, angle=-90]{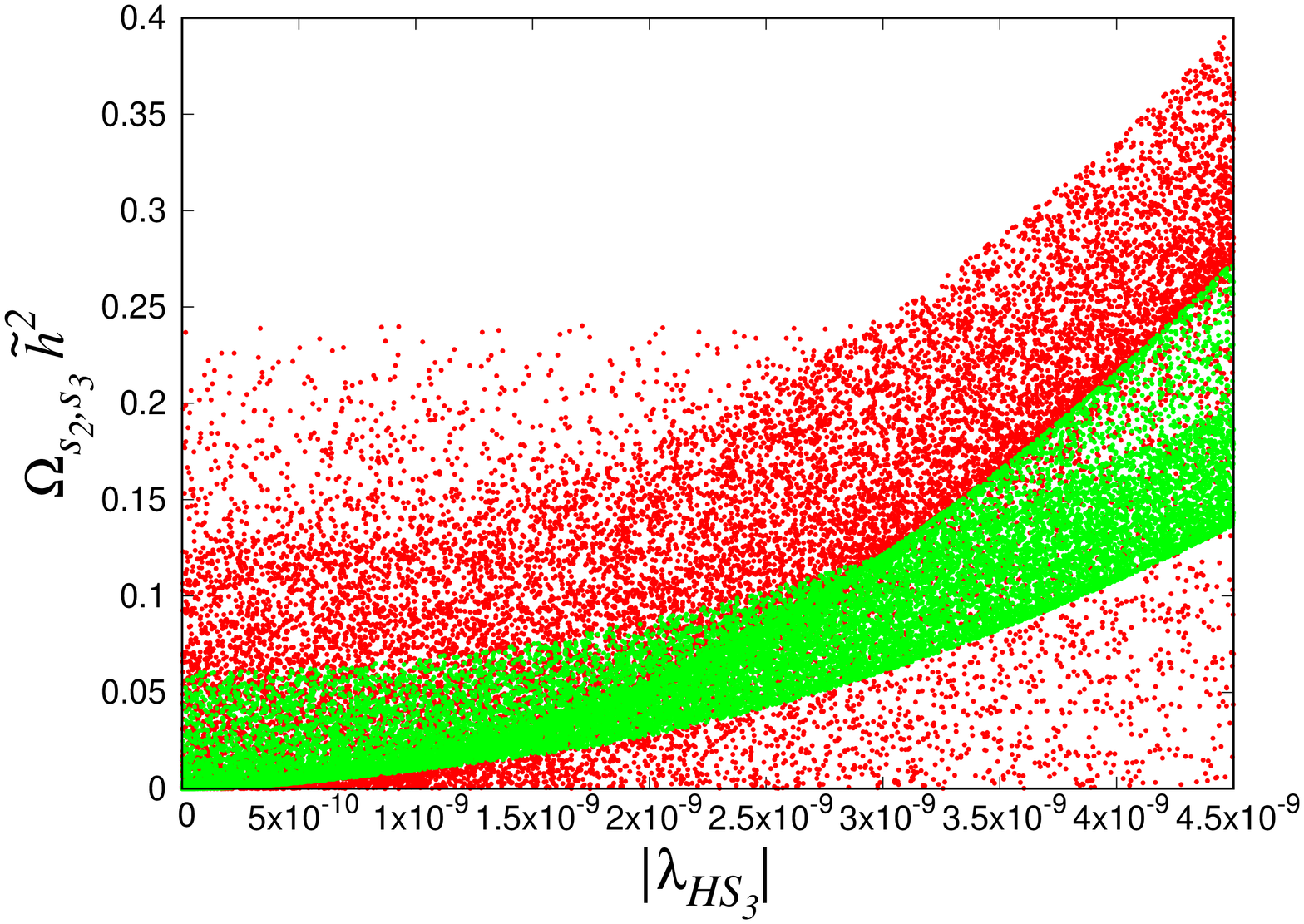}}
\subfigure[]{
\includegraphics[width=5.5 cm, height=5.5 cm, angle=-90]{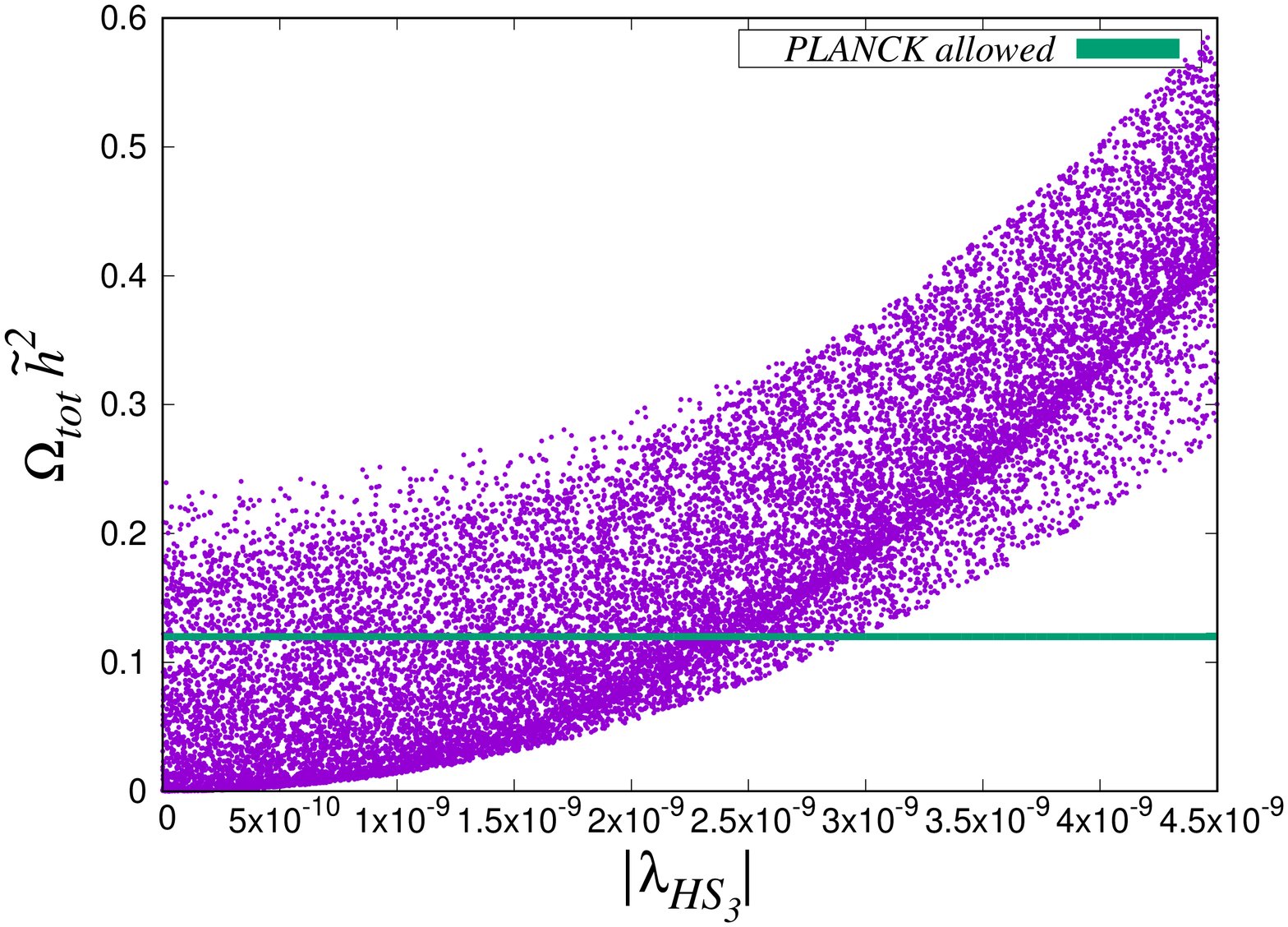}}
  \caption{{Same as Fig. 3 but for keV mass regime. See text for details.}}
   \label{b9}
\end{center}
\end{figure}

\begin{figure}[h!]
\begin{center}
\subfigure[]{
\includegraphics[width=5.5 cm, height=5.5 cm, angle=-90]{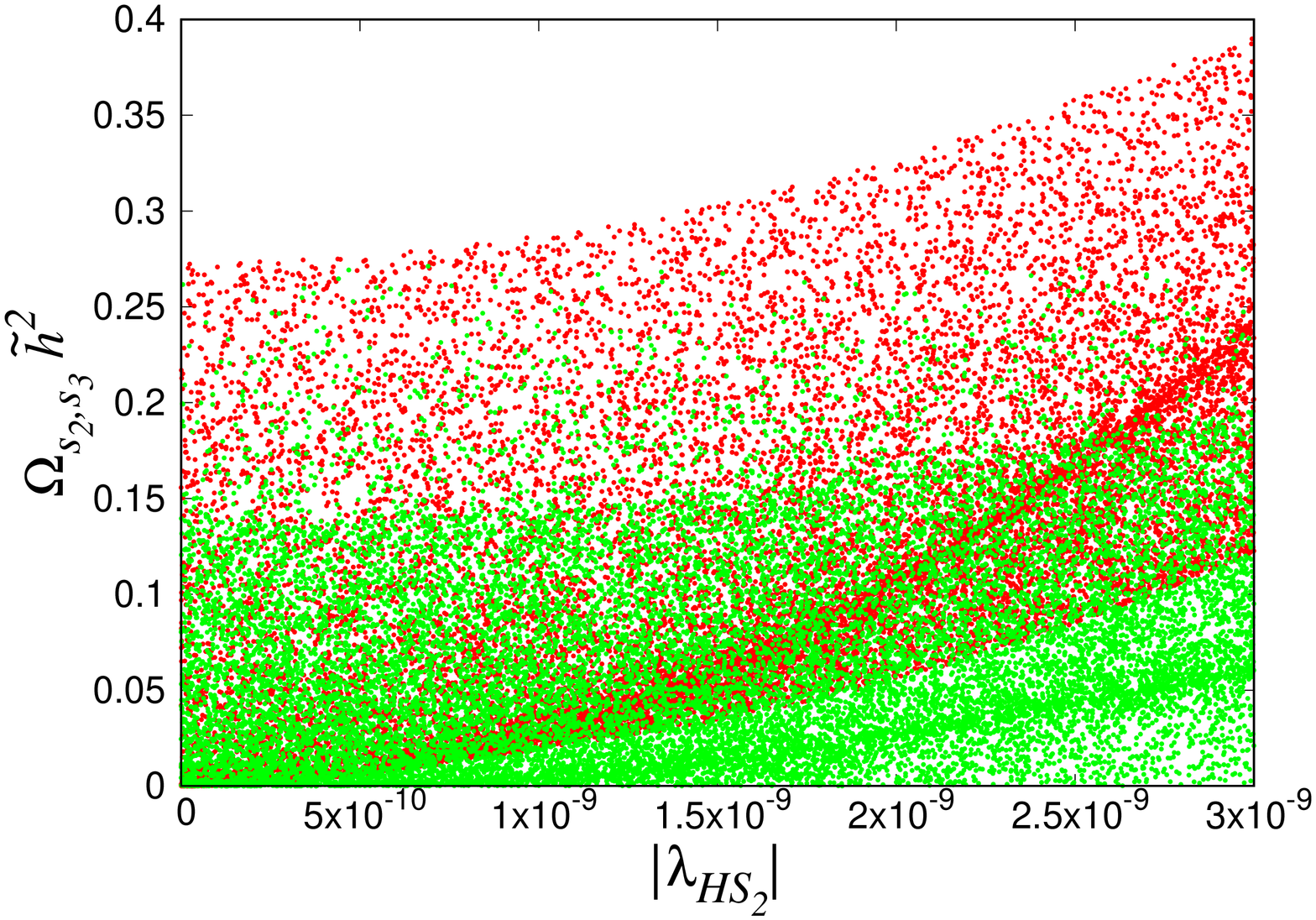}}
\subfigure[]{
\includegraphics[width=5.5 cm, height=5.5 cm, angle=-90]{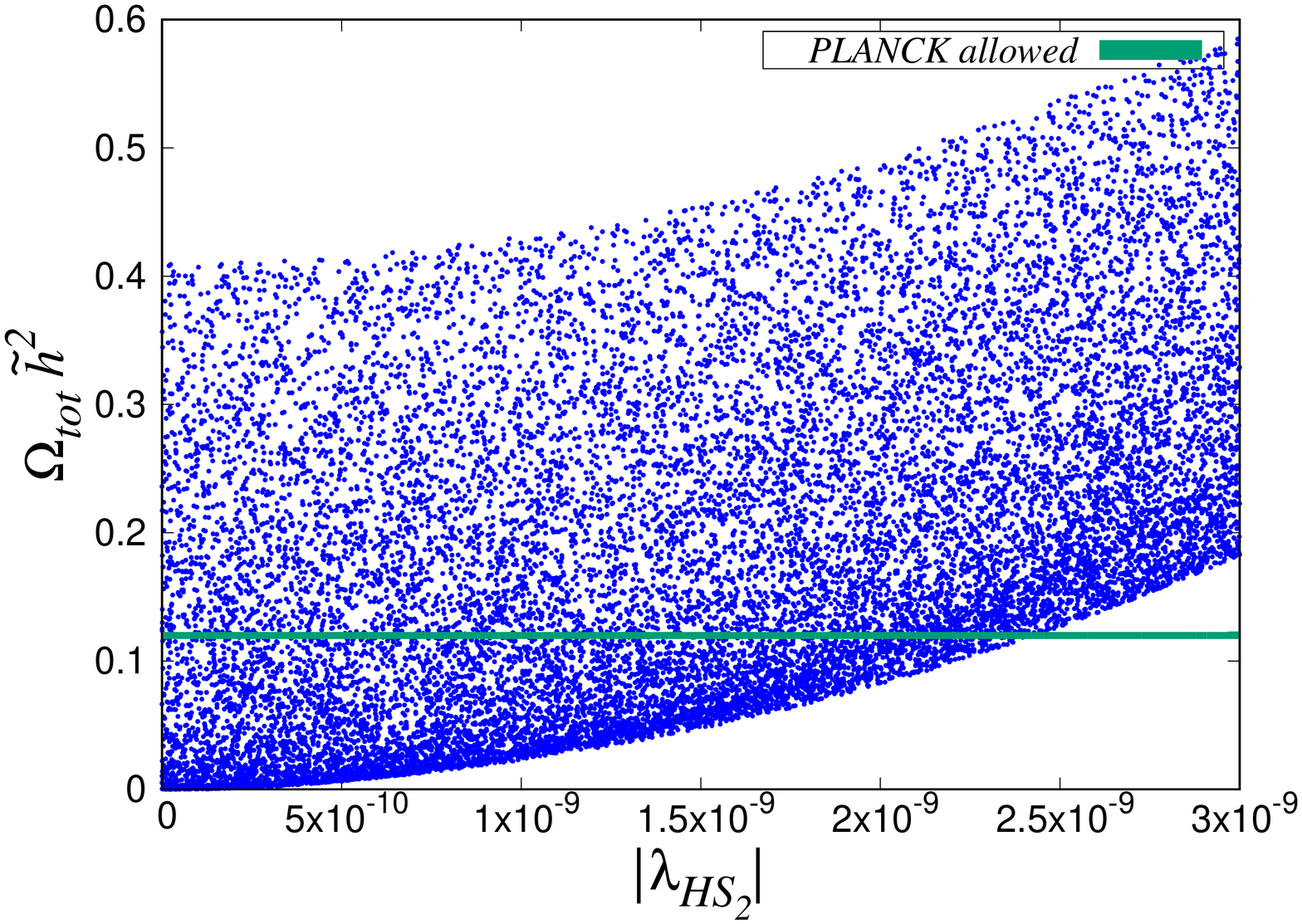}}
  \caption{{Same as Fig. 4 but for keV mass regime.}}
   \label{b10}
\end{center}
\end{figure}


\begin{figure}[h!]
\begin{center}
\subfigure[]{
\includegraphics[width=5.5 cm, height=5.5 cm, angle=-90]{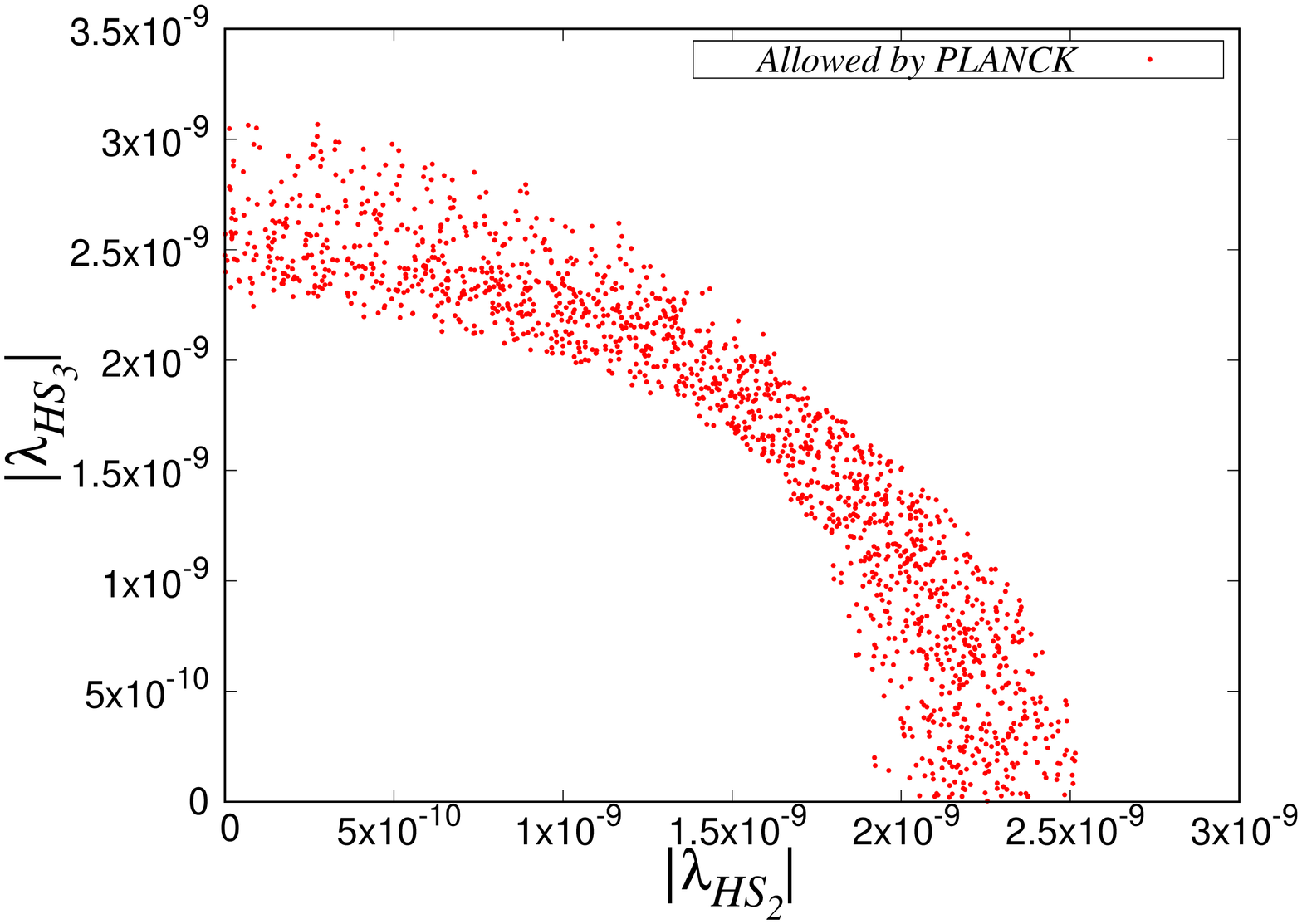}}
   \caption{{Same as Fig. 5 but for keV mass regime.}}
   \label{b12}
\end{center}
\end{figure}


%
\section{Self Interactions for Singlet Scalar Dark Matter}
Recently there are evidences of dark matter self interactions 
\cite{Harvey:2015hha,Harvey:2013tfa,Kahlhoefer:2015vua,Campbell:2015fra,Kaplinghat:2015gha}
from the 
observations  of collision of several galaxy clusters. The visible part 
of a galaxy is generally embedded inside a spherical halo of dark matter 
that extends far beyond the visible reaches of that galaxy. The dark matter 
halo makes up most of the galaxy masses. At the time of collisions between multiple galaxies a 
lareger galaxy among them pulls stars and other stellar material from a smaller galaxy and this 
process is called tidal stripping. Due to 
the presence of gravitational effect one galaxy pulls in material from another 
and this can cause the dark matter to suffer a spatial offset from the stars 
in the galaxy. Recently the galaxy cluster Abell 3827 is observed by the 
Hubble Space Telescope \cite{Kahlhoefer:2015vua}. 
The observations of the four elliptical galaxies falling into the inner 10 Kpc 
core of galaxy cluster Abell 3827 indicate that the dark matter could be self 
interacting. The position of the dark matter halos of the four falling galaxies 
can be restored by using gravitational lensing and many other strongly - lensed 
images of background objects. It is observed that one of the halos among these
four galaxies is significantly separated from its stars by a distance of 
$\Delta = 1.62^{+0.47}_{- 0.49} $ Kpc. This spatial offset can be explained by 
the study of dark matter self interaction. Determination of the size of the
spatial offset gives us an estimate of this self interaction cross-section to
the $\sigma_{\rm DM}/m \sim  1.5 \,\rm {cm}^2/\rm {g}$ which is consistent with 
the bound obtained from \cite{Harvey:2015hha}. A study on 72 colliding 
galaxy clusters \cite{Harvey:2015hha} also put an 
upper limit on the self interaction cross-section as
$\sigma_{\rm DM}/m <  0.47 \, \rm {cm}^2/\rm {g}$ with 95$\%$ C.L. 
It appears from 
\cite{Campbell:2015fra} that for the singlet scalar dark matter produced via 
thermal freeze-out 
mechanism cannot explain the observed DM self-interaction cross-section. 
The DM candidates produced via thermal freeze-in mechanism might explain 
the DM self interactions deduced from the observational results mentioned
above. In our model, as discussed earlier,
we have proposed two scalar DM candidates (two component scalar DM) 
$s_2$ and $s_3$ in FIMP scenario. 

Under the framework of present model the self interaction 
scattering cross-section per unit dark matter mass ($\sigma/m_s$) for
singlet scalar dark matter can be wriiten as \cite{Campbell:2015fra},
\bea
\frac{\sigma}{m_s} \simeq \frac{9\lambda^2} {2\pi m_s^3} \,\,\, ,
\eea
where $\lambda = \lambda_S$ for mass of dark matter to be much higher than
mass of Higgs and $\lambda = \lambda_S - {g^2 \over 8m_h^2}$ 
\cite{Campbell:2015fra}
when mass of dark matter
is less than that of Higgs. Here, $\lambda_S$ and $g$ denote 
the 4-point dark matter self-coupling
and the coupling of Higgs to the dark matter respectively. We coinsider $g \leq 2\pi$ in our work. 
Also $m_s$ and $m_h$ are the corresponding 
masses of dark matter and the Higgs.
In case of two scalar singlet model the above relation is modified and the 
effective scattering cross-section per unit effective dark matter mass can be 
expressed as,
\bea
\frac{\sigma}{m}\bigg{|}_{\rm eff} = f^2_{s_2}\frac{9\lambda_{S_2}^2} 
{2\pi m_{s_2}^3} + f^2_{s_3}\frac{9\lambda_{S_3}^2} {2\pi m_{s_3}^3} 
+ f_{s_2}f_{s_3}\frac{9\lambda_{S_2S_3}^2} {2\pi \mu_{s}^3} \,\,\, ,
\label{kpm}
\eea 
where $\lambda_{S_2}$, $\lambda_{S_3}$ denote the 4-point self couplings among 
each of 
$s_2$, $s_3$ respectively while $\lambda_{S_2S_3}$ denotes the same between 
$s_2$ and $s_3$. 
In Eq. (\ref{kpm}) $f_{s_2}$ and 
$f_{s_3}$ are respectively the corresponding dark matter density fractions
$f_i = \frac {\Omega_i} {\Omega_{\rm DM}}\,\, , i = s_2, s_3$ 
\cite{madhu,mark} for $s_2$ and $s_3$. Since $f_{s_2} +f_{s_3} = 1 
(f_{s_2} = 1- f_{s_3})$, Eq. (\ref{kpm}) reduces to the form 
\bea
\frac{\sigma}{m}\bigg{|}_{\rm eff} = f_{s_2}^2 \left(\frac{9\lambda_{S_2}^2} 
{2\pi m_{s_2}^3} + 
 \frac{9\lambda_{S_3}^2} {2\pi m_{s_3}^3} - \frac{9\lambda_{S_2S_3}^2} 
{2\pi \mu_{s}^3}\right)  
+ f_{s_2}\left(\frac{9\lambda_{S_2S_3}^2} {2\pi \mu_{s}^3} - 
2\frac{9\lambda_{S_3}^2} {2\pi m_{s_3}^3}\right)
+ \frac{9\lambda_{S_3}^2} {2\pi m_{s_3}^3}\,\,\, .
\label{kpm2}
\eea 
Using the observational bounds on $\frac{\sigma}{m}\bigg{|}_{\rm eff}$ 
one may restrict the parameter 
space $f_{s_2}-m_{s_2}-m_{s_3}$ from Eq. (\ref{kpm2}). For this purpose  upper 
bounds on the couplings  
$\lambda_{S_2}$, $\lambda_{S_3}$ and $\lambda_{S_2S_3}$ from 
perturbative unitarity conditions are used and $\frac {\sigma} {m}$ 
is calculated using Eq. (\ref {kpm2}) for a range of masses $m_{s_2}$
and $m_{s_3}$ for the two components with different fixed chosen 
values of $f_{s_2}$ ($0 < f_{s_2} < 1$). In Fig. 12 we plot for diferent 
$f_2$ values those pairs of $m_{S_2}$ and $m_{S_3}$ in 
$m_{S_2} - m_{S_3}$ in plane that satisfy the limit 
$\frac {\sigma} {m} = 0.47$. Thus, in addition to the constraints described
in Sect. 4, the self
interaction results will further constrain the masses of the dark matter 
components.
The plots in Fig. 12 show the upper bounds on the masses of dark matter 
for different fixed $f_{s_2}$ values.

\begin{figure}
\begin{center}
\includegraphics[width=7.5 cm, height=7.5 cm, angle=0]{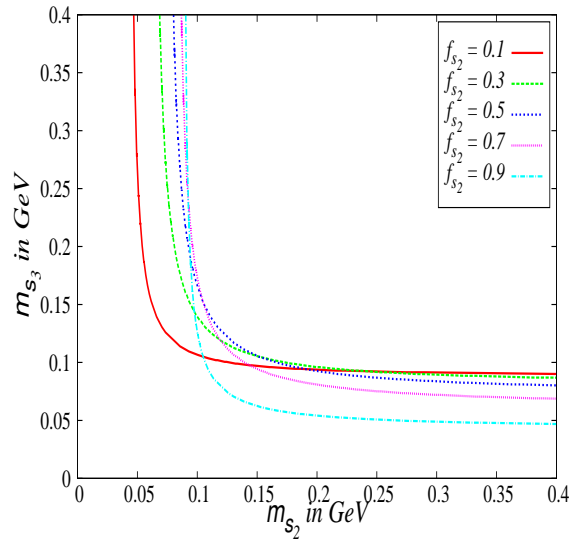}
   \caption{{The upper limits for the masses of the scalar components $m_{s_2}$ and $m_{s_3}$ that satisfy the dark matter self interaction limit (from observations) for different chosen values of fractional densities of $s_2$ component. See text for details.}}
   \label{a}
\end{center}
\end{figure}

Each pair of points on a plot in Fig. 12 
for a fixed value of 
$f_{s_2}$ therefore correspond to the upper limit of the masses for the 
components $s_2$ and
$s_3$ that satisfy the self interaction observational upper bound given in 
Eq. (\ref{kpm}). The left region of each such plot in Fig. 13 therefore 
describes the 
allowed region for the masses of $s_2$ and $s_3$ 
for a chosen fractional density 
($f_{s_2}$ and therefore $f_{s_3} = (1.0 - f_{s_2})$). From these plots of 
Fig. 12 it reveals that there are upper bounds of masses for each
of the scalar components beyond which the experimental bound for 
$\sigma/m$ will not be satisfied. Moreover, it can also be seen that 
the maximum values of $m_{s_2}$ and $m_{s_3}$ (for the chosen maximum 
value of $\frac {2\pi} {3}$ for the couplings) do not 
exceed $\sim 0.2$ GeV. These maximum limits of the individual masses 
$m_{s_2}$ or $m_{s_3}$ also vary for different fractional densities
$f_{s_2}$ ($f_{s_3}$) of the respective components.   
For example, for $f_{s_2} = 0.3$ 
($f_{s_3} = 0.7$), the mass of the $s_2$ component does not exceed a value 
$\sim 0.06$ GeV while the mass of the $s_3$ component remains limited to 
a value of around 0.15 GeV. Again, for $f_{s_2} = 0.9$ ($f_{s_3} = 0.1$) $-$ 
a situation when the two 
component dark matter is overwhelmingly dominated by only the $s_2$ 
component $-$ the upper limit for $m_{s_2} \sim 0.11$ GeV. Similar results, 
but for just one component dark matter scenario is given 
earlier by Campbell et al 
\cite{Campbell:2015fra}. Here we show, for the case of a two component scalar FIMP 
dark matter model, the simultaneous limits for the masses of the 
two components restricted by the self interaction bounds. It is also 
to be noted that although a FIMP dark matter scenario appears to be viable 
candidate in the mass range as high as few GeV from the present analysis of Sect. 6, 
such mass range is disallowed from self interaction considerations.

\section{Summary and Discussions}
The key feature of FIMP dark matter is that they were 
never in thermal equilibrium to the Universe's heat bath and are produced 
non-thermally while they approach their "freeze-in" density. 
Their couplings with SM particles are so feeble that they never attempt 
thermal equilibrium. But these types of feebly coupled dark matter
may have significance in cosmological or astrophysical contexts such as 
formation of small scale structures, signatures of the primordial 
initial conditions present in the Universe or to address issues like 
``too big to fail problem" etc.  

In this work we extend the scalar sector of Standard Model by introducing 
two singlet scalars where these scalars are considered to have 
produced in the early Universe via Feebly Interacting Massive Particle or 
FIMP mechanism. 
We perform extensive phenomenology of such a model 
and show that our two component 
FIMP scalars can be a viable candidate for dark matter 
in the Universe. Using the theoretical constraints on the interaction 
potential as well as the couplings as also employing the PLANCK 
observed relic densities and collider bounds, we domonstrate that 
in FIMP scenario, the mass regime of such scalar FIMP dark matter 
candidates may extend from GeV to keV. We have also explored 
the self interaction for these dark matter candidates. The self 
interaction cross section bound obtained from the results of 72 
colliding galaxy clusters however restricts the viable 
mass range to upper values of around $\sim 0.2$ GeV.
 
A FIMP dark matter has various cosmological and astrophysical 
implications as well as implications on its direct and indirect signatures 
\cite{1706.07442}. As the couplings of such candidates are 
extremely small it is difficult to obtain measurable direct signatures 
arising out of elastic scattering or indirect signatures from  
annihilation of FIMP dark matter. 
However, signals from decay of nonthermal light dark matter in the form 
of observed X-ray signals (3.55 keV line \cite{bulbul}) have been 
explored previously by one of the present authors \cite{anirban}.
The issues such as small scale structure formation problems can be 
addressed by warm dark matter with non thermal velocity distribution 
which is possible if they are produced via ``freeze-in" mechanism 
\cite{1706.07442}. As the FIMP dark matter never attains thermal 
equilibrium due to their feeble coupling, the initial condition for such 
non thermal production at early Universe is not washed away and can be 
probed via FIMP dark matter studies. Any primordial fluctuations caused 
by very feeble interactions of scalar fields in dark sector (which 
may not be washed away to absence of thermalisation) can be probed 
by their possible imprints in Cosmic Microwave Radiation (CMB). 

The FIMP dark matter therefore has wide implications not only 
in addressing various dark matter related issues but other 
astrophysical and cosmological concerns as well as the particle 
nature of dark matter. A two component or multicomponent dark matter
in this scenario may be useful to probe simultaneously various aspects 
related to dark matter ranging from cosmology or astrophysics to 
particle physics.

\vskip 3mm
\noindent{\Large \bf Acknowledgements}

The authors thank A. Dutta Banik and A. Biswas for their useful comments 
and suggestions. The authors would also like to thank Dr. Tommi Tenkanen for his useful suggestions. One of the authors (MP)
thanks the DST-INSPIRE fellowship grant by DST, Govt. of India.

\vskip 3mm
\noindent{\Large \bf Appendix}

The expressions for the couplings used in this work are listed below
\bea
g_{hhh} &=& - \lambda_H v \,\, ,\nonumber\\
g_{hs_2s_2} &=& - \lambda_{HS_2} v \,\, ,\nonumber\\
g_{hs_3s_3} &=& - \lambda_{HS_3} v \,\, ,\nonumber\\
g_{hhs_2s_2} &=& - \displaystyle\frac {\lambda_{HS_2}} {2} \,\, , \nonumber\\
g_{hhs_3s_3} &=& - \displaystyle\frac {\lambda_{HS_3}} {2} \,\, , \nonumber\\
g_{s_2s_2s_3s_3} &=& - \lambda_{S_2S_3} \,\, , \nonumber\\
g_{s_2s_2s_2s_2} &=& -\displaystyle\frac {\lambda_{S_2}} {4} \,\, ,\nonumber\\
g_{s_3s_3s_3s_3} &=& - \displaystyle\frac {\lambda_{S_3}} {4} \,\, , \nonumber\\
g_{WWh} &=& \displaystyle\frac {2m^2_{W}} {v}\,\, , \nonumber\\
g_{ZZh} &=& \displaystyle\frac {m^2_{Z}} {v}\,\, , \nonumber\\
g_{ffh} &=& \displaystyle\frac {m_f} {v}\,\, . \nonumber
\label{couplings}
\eea

\end{document}